\documentclass[prb,twocolumn,showpacs,preprintnumbers,amsmath,amssymb,superscriptaddress]{revtex4}

\usepackage{graphicx}
\usepackage{dcolumn}
\usepackage{bm}

\usepackage{color}
\definecolor{red}{rgb}{1,0,0} 

\begin{document}


\title{Mott insulator phases and first-order melting in $\mathrm{Bi_2Sr_2CaCu_2O_{8+\delta}}$ crystals with periodic surface holes}

\author{S. Goldberg}
\author{Y. Segev}
\author{Y. Myasoedov}
\author{I. Gutman}
\author{N. Avraham}
\author{\\M. Rappaport}
\author{E. Zeldov}
\affiliation{Department of Condensed Matter Physics, The Weizmann Institute of Science, Rehovot 76100, Israel}
\author{T. Tamegai}
\affiliation{Department of Applied Physics, The University of Tokyo, Hongo, Bunkyo-ku, Tokyo 113-8656, Japan}
\author{C. W. Hicks}
\author{K. A. Moler}
\affiliation{Department of Physics and Geballe Laboratory for Advanced Materials, Stanford University, Stanford, CA 94305, USA}

\date{\today}

\begin{abstract}
We measured the effects of periodic surface holes, created using a focused ion
beam, on the phase diagram of the vortex matter in high-$T_c$
$\mathrm{Bi_2Sr_2CaCu_2O_{8+\delta}}$ crystals. Differential magneto-optical
measurements show that the irreversibility line is shifted to higher fields and
temperatures, with respect to the pristine melting line. The irreversibility
line displays weak field dependence between integer matching fields indicating
multiple-flux-quanta pinning at holes. We find reduced equilibrium
compressibility of the vortex matter at integer matching fields, which is
strong evidence for the existence of thermodynamic Mott insulator phases.
Shaking with a transverse ac field surprisingly reveals first-order melting
that is not shifted with respect to the pristine melting line and that seems to
occur within the Mott insulator regions. This melting is understood to be the
first-order transition in the bulk of the crystal beneath the surface holes.
The transition is visible at the surface, despite the reduced vortex
compressibility in the top layer.
\end{abstract}


\maketitle

\section{Introduction}

Experimental and theoretical studies of high-$T_{c}$ materials with correlated
pinning centers have led to the discovery of many novel phases of vortex
matter, nonexistent in the pristine materials. These new phases arise from the
complex interplay between intrinsic point disorder, correlated disorder,
vortex-vortex interaction, and temperature. Correlated disorder in high-$T_c$
crystals is often introduced by heavy ion irradiation along the
crystallographic c-axis, which leads to a random distribution in the a-b
plane~\cite{prl:civ91,prb:mar91}. The effects of such random correlated
disorder on the high-$T_{c}$ phase diagram are relatively well-understood, both
theoretically~\cite{prb:nel93,revmp:bla94,prl:rad95,prl:lop04,prb:tya03,prl:das03}
and experimentally~\cite{prl:van01,prl:ban03,prl:men03,prl:ban04}. Measurements
of crystals with periodic correlated disorder are limited, since it is not yet
known how to physically realize periodic correlated disorder in thick samples.
Experimental studies of periodic disorder must therefore choose between two
options: study of thin samples, or study of artificial pins located at the
sample surface. Efforts have focused mainly on the study of thin
superconducting films with periodic pinning
centers~\cite{apl:fio73,ssc:lyk93,prl:mor98,prb:mos98,prl:bae95,prl:mar97,prb:ros96,epl:met98,prb:met99,physc:del02,prb:sil04}.
However, thin films do not necessarily retain the thermodynamic properties of
the bulk crystal, due to enhanced point disorder. Comparison of the resulting
vortex phases to the thermodynamic phases of the pristine material is thus
usually not possible.

Theoretically, the thermodynamics of vortices in the presence of \emph{random}
correlated disorder have been studied extensively. Nelson and
Vinokur~\cite{prb:nel93} mapped the pinned vortex matter onto a system of
quantum 2D bosons. They predicted the Bose glass transition from the
low-temperature Bose glass phase, in which vortices are localized, to a
higher-temperature delocalized vortex phase. They also discussed the
possibility of an incompressible Mott insulator (MI) phase, when the magnetic
induction of the sample $B$ exactly equals the matching field
$B_\phi=\rho\phi_0$, where $\rho$ is the density of the columnar defects.
Radzihovsky~\cite{prl:rad95} extended this model to include additional phases
for $B>B_{\phi}$: a low-temperature weak Bose glass phase, in which both
vortices residing at pins and those at interstitial sites are localized, an
interstitial liquid phase at intermediate temperature, in which interstitial
vortices are free to move but those at pinning sites are still pinned, and a
homogenous liquid phase at higher temperature, in which all vortices are
delocalized. For \emph{periodic} pinning centers, the various Bose glass phases
may be modified~\cite{physc:kha93}. For periodic surface holes, even such a
modified description is expected to be valid only within some finite depth from
the surface of the superconductor.

Simulations of two-dimensional systems containing periodic pinning
centers that allow only single-vortex occupancy demonstrate
commensurate states at integer matching fields $nB_{\phi}$, with
permitted values of $n$ depending on the geometry of the pinning
centers~\cite{prb:rei98}. Solutions of Ginzburg-Landau theory reveal
additional commensurate states with multi-quanta
vortices~\cite{prb:dor99,prb:ber06} for more general sample and
pinning center parameters. Different melting scenarios have been
demonstrated for triangular and kagom\'{e} arrays at low matching
fields~\cite{prb:lag01}, and for square pinning arrays both at and
in-between matching fields~\cite{prb:rei01}. The square pinning
array at the first matching field displays three phases: a
low-temperature pinned solid with square geometry, an unpinned
(``floating") solid with triangular geometry, and a high-temperature
liquid that lacks long-range order. At higher commensurate matching
fields, the floating solid phase is not found, but an intermediate
phase with mobile interstitial vortices similar to the liquid is
observed. Incommensurate fields display a pinned
phase at low temperatures with extra vortices located at interstitial positions, a phase
at intermediate temperatures in which some vortex motion is present
with both interstitials and pinned vortices participating, and a
phase at higher temperatures, in which all vortices are mobile. The
temperature at which mobility is observed for incommensurate fields
is lower than the melting temperature at the commensurate matching
fields~\cite{prb:rei01}. For the triangular and kagom\'{e}
geometries, melting at the first matching field involves a
low-temperature pinned solid and a high-temperature liquid only.
Intermediate-temperature phases, in which some or all of the
interstitial vortices are mobile, are observed at higher matching
fields~\cite{prb:lag01}. These melting transitions are expected to
be most relevant to high-$T_c$ superconductors~\cite{prb:rei01}.

Direct imaging experiments of low-$T_c$ thin films with artificial periodic
disorder have shown that highly ordered vortex states exist at integer
$nB_{\phi}$ and fractional $(p/q)B_{\phi}$ matching fields, with $n$, $q$ and
$p$ integers~\cite{sci:har96,prl:fie02}. Due to these ordered vortex states,
such films have demonstrated commensurate effects in critical
current~\cite{apl:fio73,ssc:lyk93,prl:mor98,prb:mos98},
magnetization~\cite{prl:bae95}, magneto-resistance~\cite{prl:mar97,prb:ros96},
and magnetic susceptibility
measurements~\cite{epl:met98,prb:met99,physc:del02,prb:sil04}. Possible phases
and phase transitions of the vortex matter have been inferred from these
measurements.
Enhanced flux creep rate for $B>B_{\phi}$ was thought to be evidence of a
transition from an incompressible MI state to an interstitial liquid
state~\cite{prl:bae95}. Shapiro steps in transport measurements were understood
to be a result of the coexistence of vortices pinned to artificial pinning
sites and mobile interstitial vortices~\cite{prb:van99}. The behavior of the
critical current was interpreted as evidence of two depinning energies,
corresponding to the upper boundaries of the weak Bose glass and interstitial
liquid phases~\cite{prb:ros96}. Onsets of nonzero real and imaginary parts of
the magnetic susceptibility were tentatively identified as the lower and upper
phase boundaries of an interstitial liquid phase~\cite{physc:del02}. In
addition to these states, the possibility of a saturation number $n_s>1$,
corresponding to $n_s$ vortices at each pinning site, leads to multiple-quanta
pinned vortex states, which have been observed in many
samples~\cite{prl:mor98,prb:sil04}.

There are fewer experimental data regarding the thermodynamic phases of
high-$T_c$ superconductors with periodic artificial pinning centers. The
critical current in $\mathrm{YBa_2Cu_3O_7}$ (YBCO) thin films exhibited integer
commensurate effects~\cite{apl:cas97} over a large temperature range. Scanning
Hall probe measurements indicated that trapping of $\sim15$ flux quanta is
possible for $2.5~\mu$m-diameter holes close to $T_c$ in YBCO~\cite{prb:cri05}.
Thin crystalline $\mathrm{Bi_2Sr_2CaCu_2O_{8+\delta}}$ (BSCCO) samples with
fully-penetrating periodic holes exhibited integer~\cite{physc:ooi06} and
rational~\cite{physc:ooi07} matching effects, in magneto-resistance and
transport measurements, respectively. Similar samples with surface holes also
displayed matching effects in magneto-resistance and critical
current~\cite{physc:ooi07b}. A single study using thick BSCCO samples with
surface holes displayed integer matching in local
magnetization~\cite{physc:ooi05}. In these studies of BSCCO, the matching
effects were visible in the field and temperature ranges at which the vortex
matter is known to be in a liquid state in pristine crystals. No first-order
melting step~\cite{nat:zel95} was measured in these samples, and a full
description of thermodynamic phases and transitions is lacking.

In this work, we present an investigation of a thick BSCCO crystal,
partially patterned with periodic surface holes created by a focused
ion beam, measured using differential magneto-optics
(DMO)~\cite{nat:soi00,prb:tok02,prb:avr08} accompanied by shaking
with transverse ac field~\cite{prl:wil98}. We observe step-like
behavior of the irreversibility line (IL), which may be a result of
multi-quanta pinning to holes. We see a reduction in the DMO signal
at integer matching fields, evidence of MI phases. We find a
first-order melting transition (FOT) in the patterned regions that
is not shifted with respect to the pristine melting line. This FOT
is observed even at integer matching fields, where the vortex matter
in the surface layer is essentially incompressible. We believe this
FOT to be the melting transition of the bulk of the crystal beneath
the patterned surface.

\section{Experimental details}
\begin{figure}[tb]
\includegraphics[clip,viewport=75 35 530 370,width=0.47\textwidth]{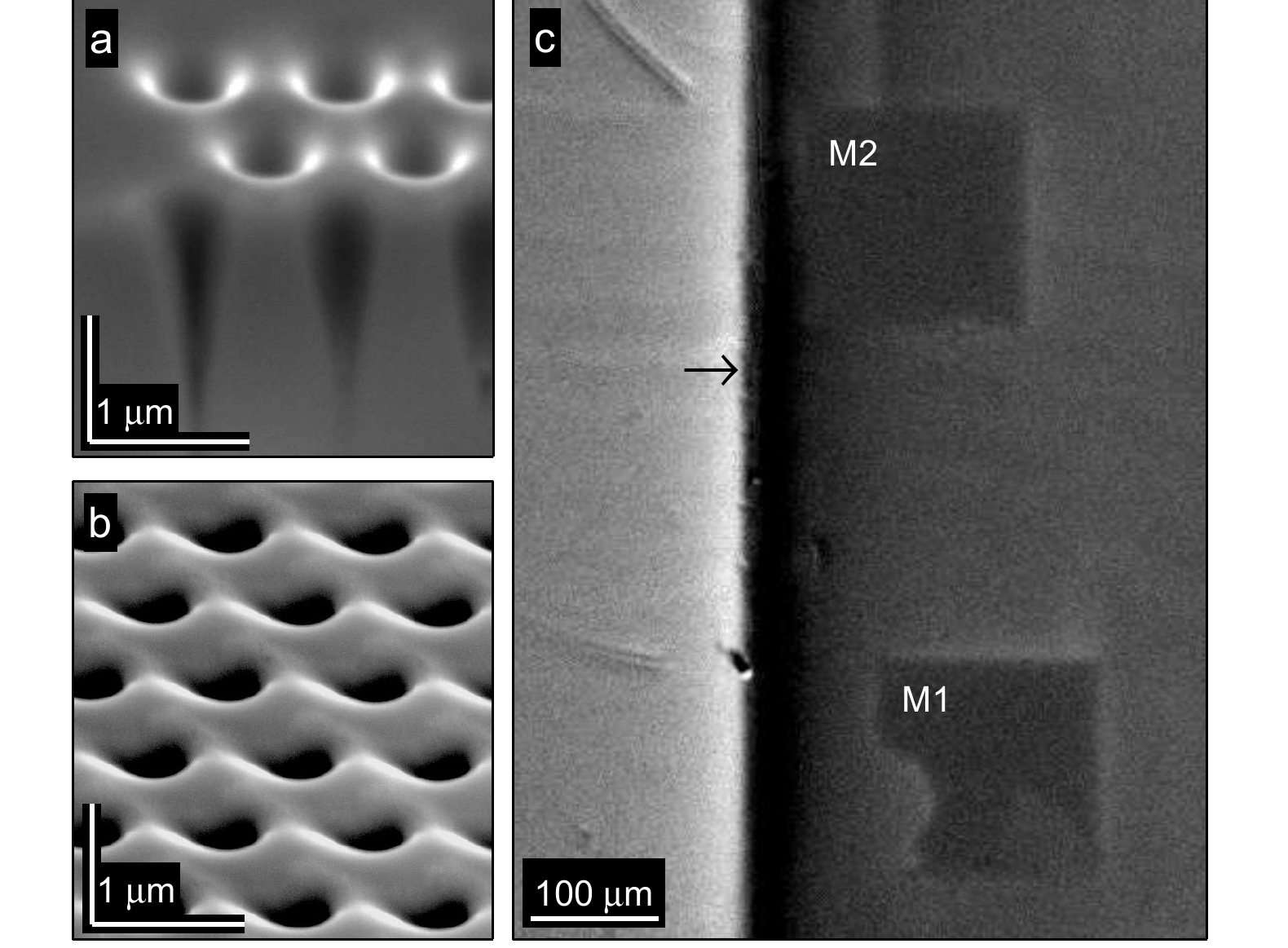}
\caption{(a) SEM image of the cross section of the surface holes.
Hole depth is $\sim1.4~\mu$m. (b) Part of one of the two arrays
patterned on the sample, imaged by SEM. The distance between holes
is $0.9~\mu$m. (c) A DMO image of the sample at $T=80~$K and
$H=21~$Oe. The sample's edge is the vertical border indicated by the
arrow. The two $B_{\phi}=29.5~\mathrm{G}$ arrays, $\mathrm{M1}$ and
$\mathrm{M2}$ of $170\times170~\mathrm{\mu m}^2$, appear as darker
regions, due to their enhanced irreversibility. The irregular shape
of $\mathrm{M1}$ is a result of surface damage that occurred after
writing the holes.}\label{fig:sample}
\end{figure}

Several samples were prepared and studied. Here we present a
detailed investigation of a $2750\times740\times30~\mu \mathrm{m}^3$
BSCCO crystal ($T_c\simeq90.5~\mathrm{K}$), with two triangular
arrays of periodic holes patterned on the top surface using an FEI
Strata 400 focused ion beam system. Figures~\ref{fig:sample}(a) and
\ref{fig:sample}(b) show SEM images of the hole profile and
periodicity, respectively. The measured hole depth was approximately
$1.4~\mathrm{\mu m}$. Hole diameter decreases from
$\sim0.6~\mathrm{\mu m}$ at the sample surface to
$\sim0.3~\mathrm{\mu m}$ at a depth of $0.7~\mathrm{\mu m}$. The
lattice constant of both arrays was $0.9~\mathrm{\mu m}$,
corresponding to a matching field of $B_{\phi}=29.5~\mathrm{G}$. The
dimensions of each array were approximately
$170\times170~\mathrm{\mu m}^2$.

DMO measurements were performed by modulating the applied field $H||$c-axis by
$\Delta H=1~$Oe while sweeping temperature $T$ at constant $H$ or scanning $H$ at
constant $T$. Each measurement point required averaging over $k$ CCD
camera exposures, first at $H+{\Delta H}/2$ and then at $H-{\Delta H}/2$, and
calculating a difference image. Each DMO image is the average of $m$ such
difference images. Using $k,m\sim10$ with a typical exposure time of $0.3~$sec
yielded a typical modulation frequency of $\sim0.33~$Hz. Values of $dB/dH$ were
derived from the DMO images by dividing the local light intensity by the intensity
of some region far from the sample, where it was assumed that
$dB/dH=1~\mathrm{G/Oe}$. For quantitative data analysis, intensities were spatially
averaged over typical area of $\sim50\times50~\mu$m$^2$. As described
previously~\cite{nat:soi00,prb:tok02,prb:avr08}, the DMO measurement with field
modulation is essentially equivalent to the measurement of the real component of
the low-frequency local ac susceptibility as obtained e.g. by Hall
sensors~\cite{prb:sch97}. Figure~\ref{fig:sample}(c) shows a DMO image of part of
the sample, taken at $T=80~\mathrm{K}$ and $H=21~\mathrm{Oe}$. The average
brightness of the patterned areas $\mathrm{M1}$ and $\mathrm{M2}$ is lower than
that of the neighboring pristine sample. This is due to an elevated irreversibility
line (IL) in the patterned areas as described below.

\section{Results}

\begin{figure}[tb]
\includegraphics[clip,viewport=1 1 500 430,width=0.47\textwidth]{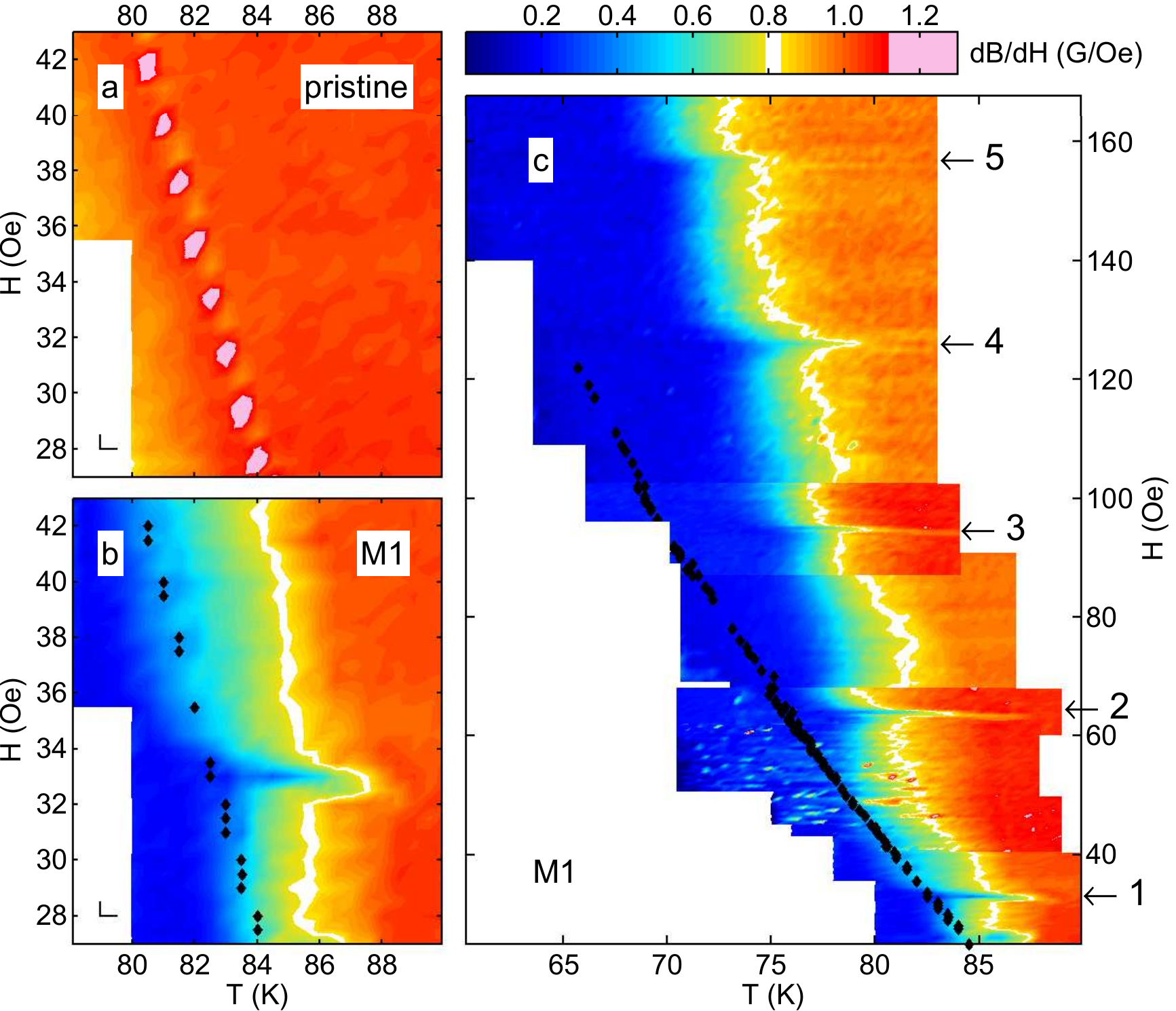}
\caption{(Color online) $dB/dH$, the change in magnetic induction
due to field modulation of $\Delta H=1~\mathrm{Oe}$, for different
sample regions, taken during $T$-scans. (a) A pristine region. The
pristine melting line $T_m$ appears as a series of light gray (light
pink) paramagnetic peaks in $dB/dH$ with values above
$1~\mathrm{G/Oe}$. This particular run was carried out on a sparse
grid in $T$ and $H$ of $0.4~$K and $0.5~$Oe (denoted by the lines in
the lower left corner). The apparent discontinuity in $T_m$ is an
artifact resulting from grid spacing being larger than the width of
the melting peak. (b) $dB/dH$ of region $\mathrm{M1}$ in the
vicinity of the first matching field, $B_{\phi}=29.5~$G. $dB/dH$
values are reduced compared to (a) and a narrow dip appears at
$H=33~$Oe. The location of the pristine melting line $T_m$ is
denoted by black points. (c) $dB/dH$ in the patterned region
$\mathrm{M1}$ over a wide range of $T$ and $H$. Matching effects
(denoted by arrows) are visible at $H=33,64,95,126,$ and $157~$Oe,
consistent with integer multiples of the predicted
$B_{\phi}=29.5~$G. The pristine melting line $T_m$ is denoted by
black points. The patches in the data are a result of slightly
differing setup parameters for the different experimental runs.}
\label{fig:PD}
\end{figure}

We first inspect the IL of the patterned regions. The IL is
important in the context of possible Bose glass phases because it is
thought to be the dynamic manifestation of the thermodynamic Bose
glass transition~\cite{prb:nel93}. In DMO measurements,
reversibility of the vortex matter is quantified by modulating the
applied field by $\Delta H$, and measuring $dB/dH$, the change in
the local magnetic induction due to the modulation. Strong pinning
results in $dB/dH=0$, whereas full reversibility corresponds to
$dB/dH=1~\mathrm{G/Oe}$. The irreversibility threshold in the
following was chosen arbitrarily at $dB/dH=0.8~\mathrm{G/Oe}$, with
$T_{IL}$ ($H_{IL}$) denoting the temperature (external field) at
which this threshold is reached. We emphasize that the resulting IL
reflects the response of the vortex system at low frequencies.
Transport measurements or DMO with current modulation could possibly
map out additional boundary lines similar to the delocalization line
of vortices from columnar defects~\cite{prl:ban04}, however, such
measurements are beyond the scope of the present study.
Figure~\ref{fig:PD} shows $dB/dH$ for the pristine region
(Fig.~\ref{fig:PD}(a)) and patterned region $\mathrm{M1}$
(Figs.~\ref{fig:PD}(b) and \ref{fig:PD}(c)), measured by $T$ scans
at constant $H$.

Focusing on the pristine region (Fig.~\ref{fig:PD}(a)), we see a
series of sharp peaks in $dB/dH$ with paramagnetic $dB/dH>1$ (light
gray, light pink online), corresponding to the first-order melting
transition $T_m$ from a low-temperature vortex solid to a
high-temperature vortex liquid~\cite{prb:mor96,nat:soi00}. The black
dots in Fig.~\ref{fig:PD}(b) show the pristine melting line $T_m$
extracted from Fig.~\ref{fig:PD}(a). The patterned region
$\mathrm{M1}$ (Fig.~\ref{fig:PD}(b)), in comparison, shows no FOT.
It does, however, exhibit two notable features. First, the IL of
region $\mathrm{M1}$ is shifted to higher temperature. This is seen
by focusing on $T_{IL}$ (white contour) in Figs.~\ref{fig:PD}(a) and
\ref{fig:PD}(b). $T_{IL}$ of the patterned region $\mathrm{M1}$ is
significantly greater than $T_{IL}$ of the pristine region, which
occurs at temperatures lower than the pristine $T_m$. The second
notable feature in Fig~\ref{fig:PD}(b) is a narrow finger near
$H=33~$Oe, approximately $1~\mathrm{Oe}$ wide, for $82<T<87~$K, in
which $dB/dH$ of $\mathrm{M1}$ is suppressed.

Figure~\ref{fig:PD}(c) shows $dB/dH$ of patterned region
$\mathrm{M1}$ over a larger range of $H$ and $T$. The pristine
melting line $T_m$ is plotted as black dots for comparison. The IL
of $\mathrm{M1}$ is clearly shifted to higher temperatures. Sharp
fingers, or narrow regions of $H$ in which $dB/dH$ of $\mathrm{M1}$
is highly suppressed, occur at $H=33,64,95,126,$ and $157~$Oe
(denoted by arrows), consistent with integer multiples
$B/B_{\phi}=1,2,3,4,$ and $5$ of the predicted matching field
$B_{\phi}=29.5~$G. The minima in $dB/dH$ as a function of $H$ at
matching fields are of both dynamic and thermodynamic origin.
Enhanced pinning at matching fields suppresses vortex motion, and
hence also $dB/dH$. As discussed below, these minima also indicate
narrow ranges of $H$ with reduced equilibrium compressibility, since
the compressional modulus $c_{11}$ is proportional to
$dH/dB$~~\cite{prb:wen98}. In addition to the fingers observed
\emph{at} integer $B/B_{\phi}$, we see that $T_{IL}$ \emph{between}
matching fields is step-like, with $T_{IL}(H)$ weakly dependent on
$H$ between matching fields, and shifts in $T_{IL}(H)$ occurring at
matching fields. For example, $T_{IL}\approx76.5~$K for
$3<H/B_{\phi}<4$, and $T_{IL}\approx74~$K for $4<H/B_{\phi}<5$
(white contour).

\begin{figure}[tb]
\includegraphics[clip,viewport=71 348 538 551,width=0.47\textwidth]{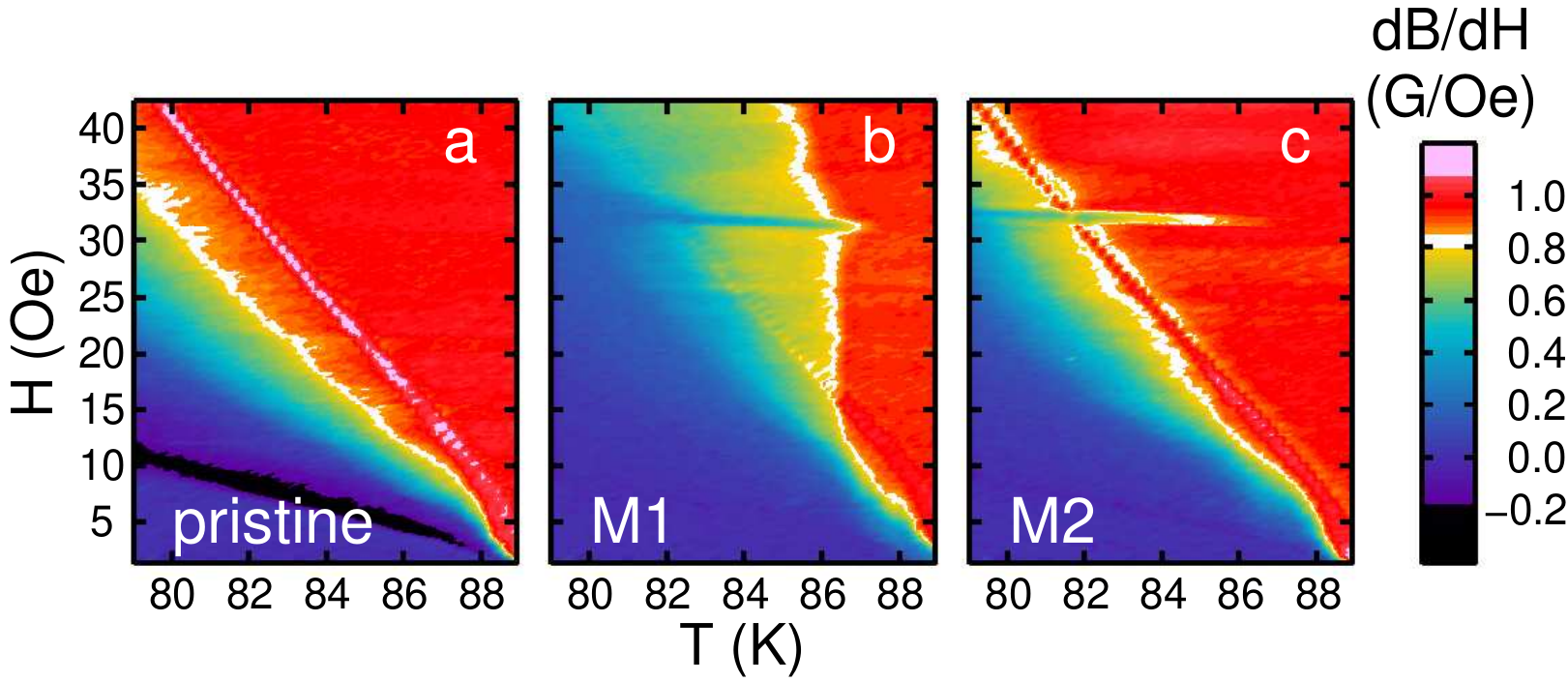}
\caption{(Color online) $dB/dH$ for $B\lesssim B_{\phi}$, measured
during $T$-scans. (a) $dB/dH$ of a pristine region. The melting
transition $T_m$ appears as a line with $dB/dH>1~\mathrm{G/Oe}$
(light gray, light pink online). The temperature $T_{IL}$ at which
the IL is located ($dB/dH=0.8~\mathrm{G/Oe}$, white) is found below
$T_m$. (b) and (c) show $dB/dH$ of patterned regions $\mathrm{M1}$
and $\mathrm{M2}$, respectively. $T_{IL}$ of $\mathrm{M1}$ and
$\mathrm{M2}$ is shifted to higher temperatures relative to the
pristine $T_{IL}$. A sharp finger in $T_{IL}$ of both $\mathrm{M1}$
and $\mathrm{M2}$ appears at $H=33~$Oe, where $B=B_{\phi}$. Negative
values of $dB/dH$ (black) correspond to negative permeability, due
to geometrical barriers.} \label{fig:lowH}
\end{figure}

We now focus on the IL of the patterned regions below and in the
vicinity of the first matching field. Figure~\ref{fig:lowH} shows
$dB/dH$ measured simultaneously in the pristine region and in the
patterned regions $\mathrm{M1}$ and $\mathrm{M2}$, for $B\lesssim
B_{\phi}$. $T_{IL}$ ($dB/dH=0.8~\mathrm{G/Oe}$, white contour) of
the pristine region is located below the pristine melting line
$T_m$. In region $\mathrm{M1}$ (Fig.~\ref{fig:lowH}(b)), $T_{IL}$ is
shifted to higher temperatures at all fields when compared to the
$T_{IL}$ of the pristine region (Fig.~\ref{fig:lowH}(a)). An
additional sharp finger in the IL, extending to $T_{IL}\simeq87~$K,
occurs at $H=33~$Oe, or $B=B_{\phi}$. The IL of patterned region
$\mathrm{M2}$ (Fig.~\ref{fig:lowH}(c)) is shifted somewhat less, yet
it too displays a sharp shift to higher $T$ at the matching field
$B_{\phi}$. The reason for the difference in the shift of the IL of
the two arrays may be a result of the difference in the arrays'
locations ($\mathrm{M2}$ is closer to the sample's edge) or due to
some difference in the holes of the two arrays, which are not
identical and may have different pinning properties. Still, for both
arrays the IL is shifted upward, with an additional sharp finger at
$B_{\phi}$. In BSCCO crystals irradiated with low concentration of
CDs the sharp finger is absent. Instead, a kink is observed in the
vicinity of $B_{\phi}$ which is believed to be the result of
depinning of two different vortex populations. Below $B_{\phi}$,
$T_{IL}$ is the temperature at which vortices located at CDs depin.
Above $B_{\phi}$, $T_{IL}$ is the depinning temperature of
interstitial vortices~\cite{prl:ban04}. Also, fractional matching
features have been observed~\cite{physc:ooi07b} in BSCCO samples
with periodic surface defects. We do not detect any fractional
matching features in the IL of the patterned regions; the reason for
this is not clear, but is consistent with results shown
elsewhere~\cite{physc:ooi05}.


The IL in BSCCO is known to be a dynamic feature of the phase
diagram~\cite{revmp:bla94,prl:maj95,nat:avr01}. However, since it
indicates a region of the phase diagram in which there is a change
in the system's dynamic response, it may indicate an underlying
thermodynamic transition that occurs at similar values of field and
temperature. In order to detect a possible underlying FOT, and to
determine whether the reduced $dB/dH$ regions are truly a
thermodynamic feature of the phase diagram, we studied the behavior
of the first matching field in the presence of shaking. The shaking
technique~\cite{prl:wil98} utilizes an in-plane, ac magnetic field.
It is known to suppress hysteretic behavior in BSCCO, enabling the
observation of thermodynamic
properties~\cite{nat:avr01,prl:bei05,prl:bei07}.

\begin{figure}[tb]
\includegraphics[clip,viewport=1 1 367 448,width=0.4\textwidth]{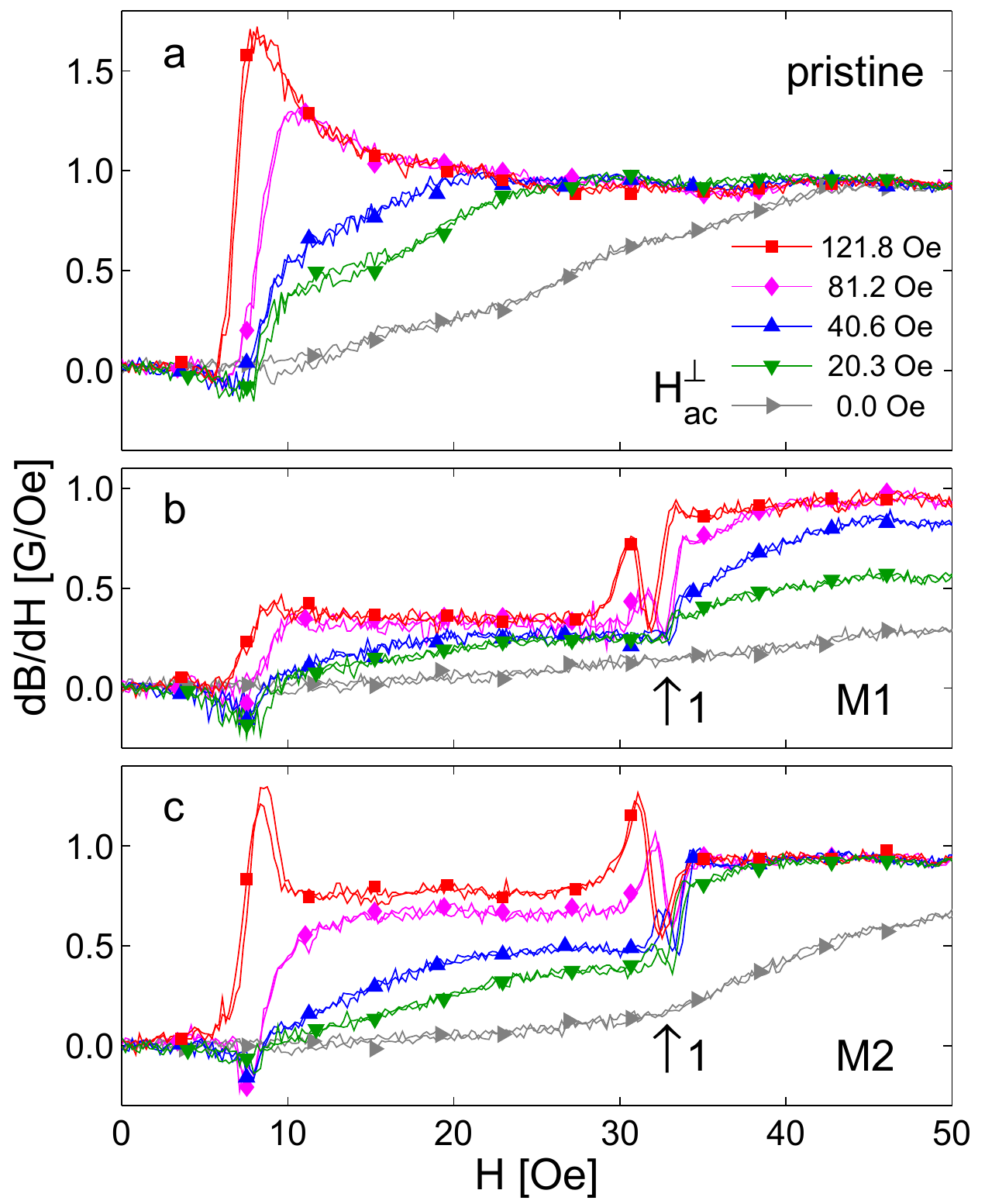}
\caption{(Color online) $dB/dH$ vs. applied field $H$ (scanned up and down) for
different values of in-plane shaking amplitude $H_{ac}^{\perp}$, at
$T=77~\mathrm{K}$ for the pristine region (a) and patterned regions $\mathrm{M1}$
(b) and $\mathrm{M2}$ (c). Arrows denote the first matching field. Data are shown
for different values of applied $H_{ac}^{\perp}$ (from bottom to top): 0 ($\rhd$),
20.3 ($\nabla$), 40.6 ($\triangle$), 81.2 ($\diamondsuit$), and 121.8 ($\Box$)~Oe.
Shaking frequency was $15~$Hz for all measurements. Symbols appear every $37$ data
points.}\label{fig:shakingEffect}
\end{figure}

The effect of shaking is shown in Fig.~\ref{fig:shakingEffect}, for
the pristine and patterned regions $\mathrm{M1}$ and $\mathrm{M2}$
at $77~$K. Increasing shaking amplitude $H_{ac}^{\perp}$ leads to a
systematic increase in $dB/dH$, which saturates near
$1~\mathrm{G/Oe}$, as expected for a fully penetrable sample. A
feature common to all three plots in Fig.~\ref{fig:shakingEffect},
and thus unrelated to the surface holes, is an abrupt change from
zero to negative $dB/dH$ at $\simeq8~$Oe. Below $\simeq8~$Oe, the
sample is in the Meissner phase, with $B=0$. $dB/dH<0$ immediately
above the Meissner phase corresponds to negative local
permeability~\cite{prl:mor96}. This effect, which occurs in BSCCO
samples with platelet geometry, is a result of the geometrical
barrier~\cite{prl:sch94,prl:zel94} and the modulation of the vortex
dome during the modulation cycle of the applied field $H\pm \Delta
H/2$. This negative permeability is also visible as a black strip at
low fields in Fig.~\ref{fig:lowH}(a). It is interesting to note that
the negative $dB/dH$ values are not visible at the highest shaking
amplitude, $H_{ac}^{\perp}=121.8~$Oe. This indicates that the
shaking field enables the vortices to overcome the geometrical
barrier. Consequently, the local negative permeability changes to
high positive local permeability, as seen in
Figs.~\ref{fig:shakingEffect}(a) and \ref{fig:shakingEffect}(c).
This is the expected behavior in the absence of geometrical
barriers, as demonstrated for prism-shaped
samples~\cite{physc:mor97}.

There are two notable differences between the pristine
(Fig.~\ref{fig:shakingEffect}(a)) and patterned
(Figs.~\ref{fig:shakingEffect}(b) and \ref{fig:shakingEffect}(c))
regions. The first difference is the appearance of the matching
feature in the form of a dip in $dB/dH$ near $32~$Oe, denoted by
arrows in Fig.~\ref{fig:shakingEffect}. Shaking extends the range of
temperatures for which this dip is visible well into the vortex
solid region below $T_m$. Without shaking, the first matching
feature is not visible at this temperature ($77~$K, see
Fig.~\ref{fig:PD}), due to the enhanced pinning in the vortex solid.
With increased $H_{ac}^{\perp}$, the matching feature appears first
as a step (Fig.~\ref{fig:shakingEffect}(b), $H_{ac}^{\perp}=20.3$
and $40.6~$Oe) and then as a dip in $dB/dH$
($H_{ac}^{\perp}=81.2~$Oe). In some cases, as shown in
Fig.~\ref{fig:shakingEffect}(c) for high $H_{ac}^{\perp}$, a peak
appears in $dB/dH$ immediately before the dip. The slight downward
shift in $H$ of the dip for increasing $H_{ac}^{\perp}$ probably
results from the increased penetration of magnetic induction $B$ at
higher $H_{ac}^{\perp}$, resulting in the same $B=B_{\phi}$ at
slightly lower values of applied field $H$. The dip in $dB/dH$
corresponds to a reduction in the compressibility of the vortex
matter at $B_{\phi}$.

\begin{figure}[tb]
\includegraphics[clip,viewport=30 40 560 420,width=0.49\textwidth]{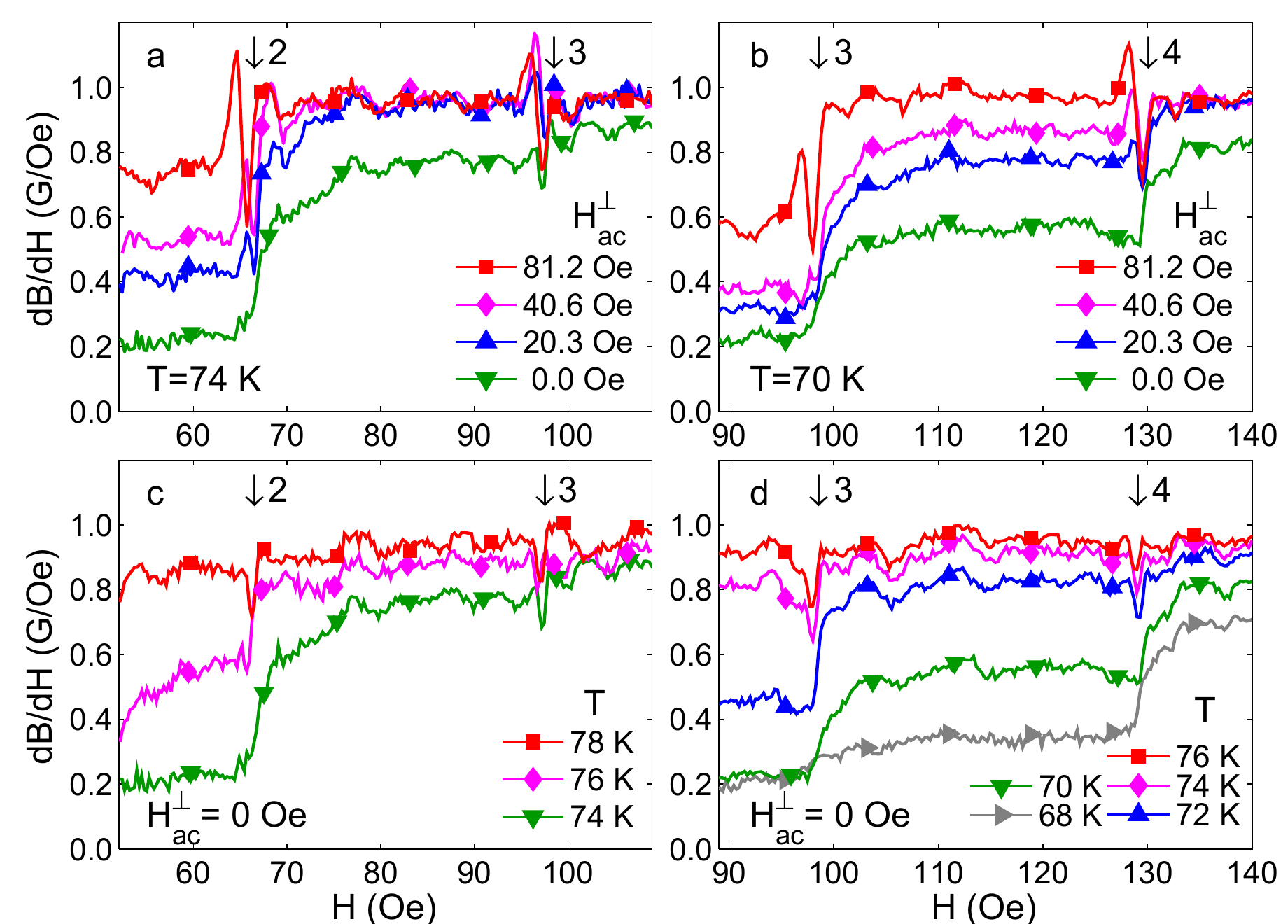}
\caption{(Color online) $dB/dH$ of region $\mathrm{M2}$ as a function of $H$
for different temperatures $T$ and shaking amplitudes $H_{ac}^{\perp}$. The
effect of shaking is similar to the effect of temperature (see text). (a)
Matching effects at $T=74~$K, $B/B_{\phi}=2,3$, and (b) at $T=70~$K,
$B/B_{\phi}=3,4$. $H_{ac}^{\perp}=0$ ($\nabla$), $20.3$ ($\triangle$), $40.6$
($\diamondsuit$), and $81.2$ $(\Box)~$Oe. (c) Matching effects at $T=74$
($\triangle$), $76$ ($\diamondsuit$) and $78~(\Box)~$K, $B/B_{\phi}=2,3$,
without shaking. (d) Matching effects at $T=68$ ($\rhd$), $70$ ($\nabla$), $72$
($\triangle$), $74$ ($\diamondsuit$), and $76~(\Box)~$K, $B/B_{\phi}=3,4$,
without shaking. Arrows denote matching fields. Symbols appear every $30$ data
points.}\label{fig:higherOrders}
\end{figure}

The second difference between the pristine and patterned regions in
Fig.~\ref{fig:shakingEffect} can be seen away from the matching field. For the
pristine region, the values of $dB/dH$ increase gradually from $dB/dH\simeq0$
at low field to $dB/dH=1~\mathrm{G/Oe}$ at sufficiently high applied field $H$.
For the patterned regions, the behavior of $dB/dH$ is plateau-like, with the
matching feature dividing between neighboring plateaus. This can be seen in
Fig.~\ref{fig:shakingEffect}(b) for $H_{ac}^{\perp}=81.2,121.8~$Oe, and in
Fig.~\ref{fig:shakingEffect}(c) for $H_{ac}^{\perp}=40.6~$Oe. These plateaus
are consistent with the observed step-like behavior of $T_{IL}$ during
$T$-scans, as shown in Fig.~\ref{fig:PD}(c). The plateaus in $dB/dH$ appear
between matching fields, where $T_{IL}$ is almost independent of $H$. The step
between the plateaus appears at $H=B_{\phi}$, consistent with the steps in
$T_{IL}$ that occur at $H=nB_{\phi}$. The value of $dB/dH$ for each plateau in
Figs.~\ref{fig:shakingEffect}(b) and ~\ref{fig:shakingEffect}(c) increases with
increasing $H_{ac}^{\perp}$. Figures~\ref{fig:higherOrders}(a) and
\ref{fig:higherOrders}(b) show the effects of shaking with different values of
$H_{ac}^{\perp}$ for $B/B_{\phi}=2,3$ and $B/B_{\phi}=3,4$, respectively, for
patterned region $\mathrm{M2}$. The data for $\mathrm{M1}$ are similar. Clearly
the same matching features that appear for $B/B_{\phi}=1$, namely a step in
$dB/dH$ for low $H_{ac}^{\perp}$ that develops into a dip for higher
$H_{ac}^{\perp}$, are visible also for higher matching fields.

Interestingly, increasing $T$ and increasing shaking amplitude
$H_{ac}^{\perp}$ have similar effects on $dB/dH$ immediately below
the IL. This can be seen by comparing
Figs.~\ref{fig:higherOrders}(a) and \ref{fig:higherOrders}(c), in
which $dB/dH$ of region $\mathrm{M2}$ in the vicinity of
$B/B_{\phi}=2,3$ is plotted for different values of $H_{ac}^{\perp}$
and $T$, respectively. Increasing $H_{ac}^{\perp}$ and increasing
$T$ (bottom to top curves) both tend to increase $dB/dH$, and both
have the effect of transforming the matching feature from a step to
a dip. However, the overshoot in $dB/dH$ immediately below the
matching feature appears only at nonzero $H_{ac}^{\perp}$. Similar
behavior is observed for $B/B_{\phi}=3,4$ in
Figs.~\ref{fig:higherOrders}(b) and \ref{fig:higherOrders}(d).


\begin{figure}[t!]
\includegraphics[clip,viewport=34 304 591 619,width=0.48\textwidth]{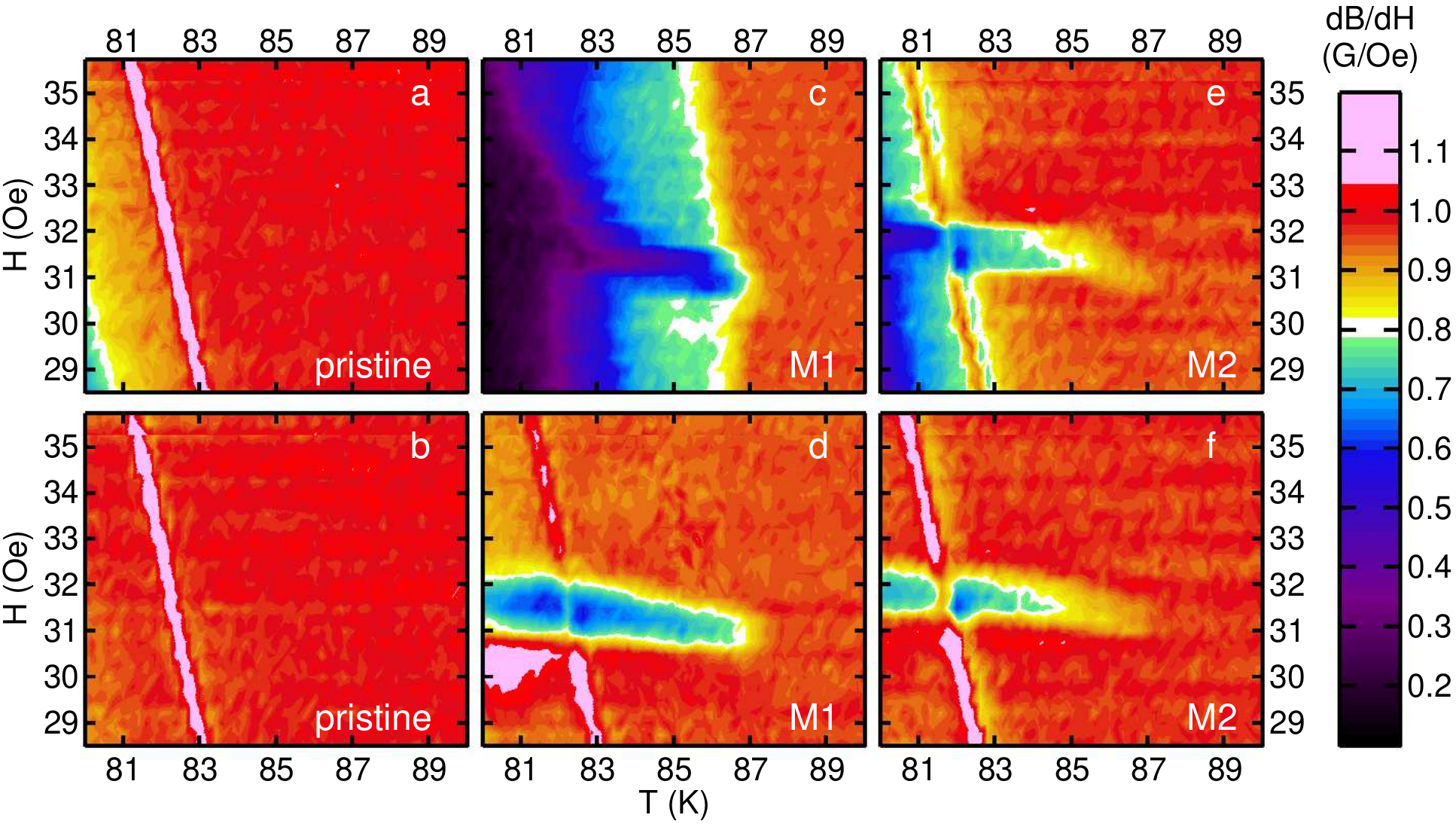}
\caption{(Color online) $dB/dH$ as a function of $H$ and $T$ near the
intersection of melting and $B/B_{\phi}=1$ matching, without (top panels) and
with (bottom panels) shaking. Shaking parameters were $15~$Hz and $81.2~$Oe.
Results are shown for the pristine sample ((a),(b)) and patterned regions
$\mathrm{M1}$ and $\mathrm{M2}$ ((c),(d) and (e),(f), respectively).}
\label{fig:meltMatch}
\end{figure}

\begin{figure}[tb]
\includegraphics[clip,viewport=34 304 591 619,width=0.48\textwidth]{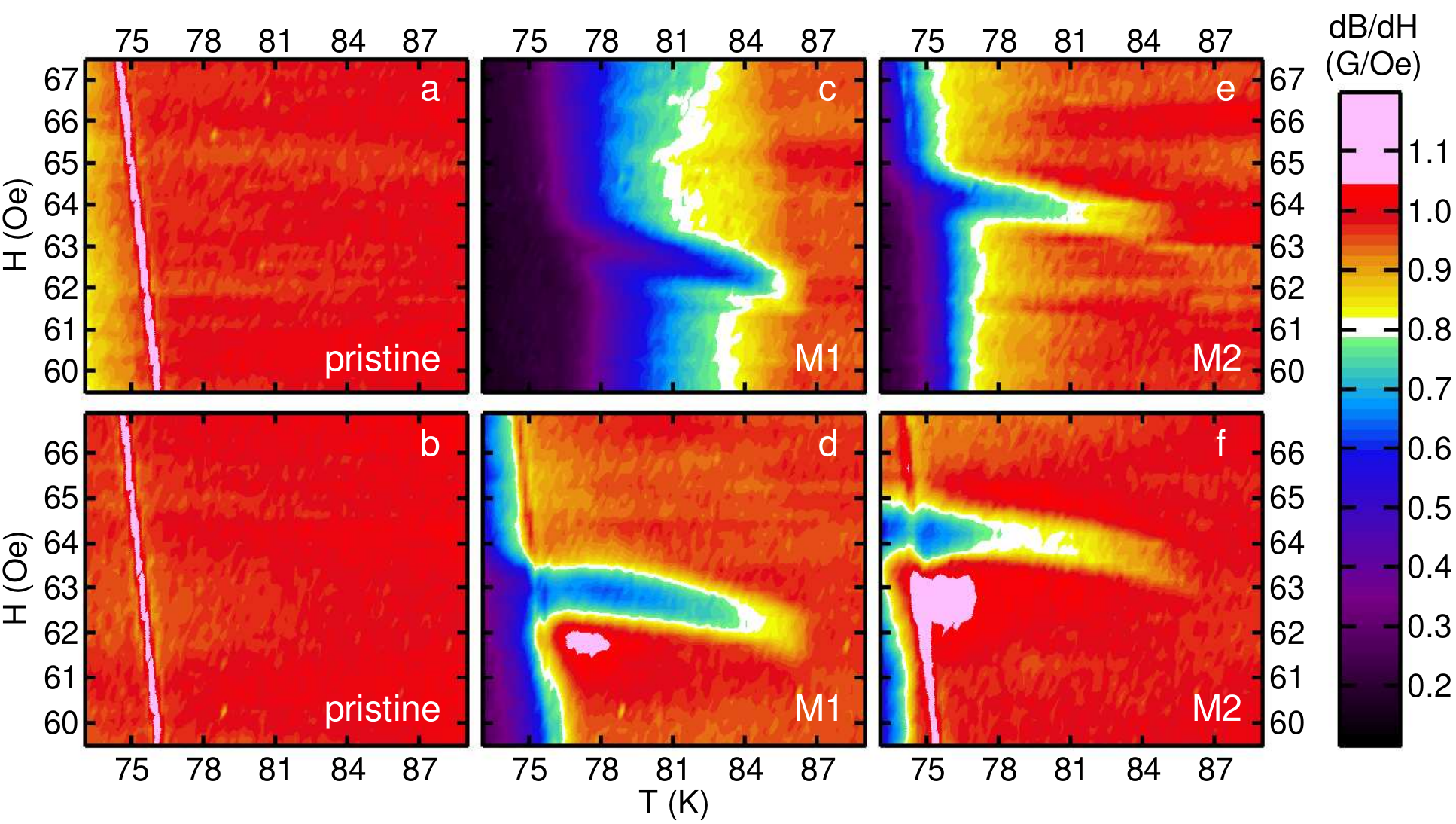}
\caption{(Color online) $dB/dH$ as a function of $H$ and $T$ near the
intersection of melting and $B/B_{\phi}=2$ matching, without (top panels) and
with (bottom panels) shaking. Shaking parameters were $15~$Hz and $81.2~$Oe.
Results are shown for the pristine sample ((a),(b)) and patterned regions
$\mathrm{M1}$ and $\mathrm{M2}$ ((c),(d) and (e),(f), respectively). The slight
curving of the matching effect to lower $H$ for higher $T$ in (c),(d),(e) and
(f) is due to the increased penetration $B(H)$ at higher $T$.}
\label{fig:meltMatch2}
\end{figure}

We now address the question of first-order melting within the
patterned regions in the presence of shaking. For the results shown
below, we applied a $15~\mathrm{Hz}$, $H_{ac}^{\perp}=81.2~$Oe
shaking field. We find that shaking shifts the IL to lower fields
and temperatures and thus enables the observation of a FOT in the
patterned regions of the sample. Figure~\ref{fig:meltMatch} shows
detailed scans of the $H-T$ region in which the pristine $T_m$ line
intersects the $B/B_{\phi}=1$ matching line. $dB/dH$ is shown for
both patterned regions and for the pristine region, without and with
shaking (top and bottom panels, respectively). For the pristine
region (Figs.~\ref{fig:meltMatch}(a) and ~\ref{fig:meltMatch}(b)),
$dB/dH$ at lower $T$ and $H$ is raised slightly by shaking, and the
pristine melting line $T_m$, which appears as a line with
paramagnetic $dB/dH>1~\mathrm{G/Oe}$ (light gray, light pink
online), remains essentially unchanged. For the patterned region
$\mathrm{M1}$, no FOT was visible without shaking
(Fig.~\ref{fig:meltMatch}(c)). With shaking
(Fig.~\ref{fig:meltMatch}(d)), a FOT became visible. It appears to
be located at the same temperatures and fields as the pristine
melting line $T_m$. Remarkably, the $T_m$ line is clearly visible
even at the bottom of the $B_{\phi}$ matching dip. For the patterned
region $\mathrm{M2}$, shaking was not needed to uncover the FOT
(Fig.~\ref{fig:meltMatch}(e)). While shaking raised $dB/dH$ values
overall, it did not change the location or the nature of the FOT
(Fig.~\ref{fig:meltMatch}(f)). Similar results are shown for
$B/B_{\phi}=2$ in Fig.~\ref{fig:meltMatch2}. In this case, shaking
was necessary to view the FOT in both patterned regions
$\mathrm{M1}$ (Fig.~\ref{fig:meltMatch2}(c) and
\ref{fig:meltMatch2}(d)) and $\mathrm{M2}$
(Fig.~\ref{fig:meltMatch2}(e) and \ref{fig:meltMatch2}(f)). Note
that the location of the FOT of the patterned regions in the phase
diagram is indistinguishable from the location of the pristine
melting line, $T_m$. Moreover, $T_m$ and the $nB_{\phi}$ lines seem
to intersect with no apparent interaction, as if the periodic
pinning potential of the holes has no effect on melting. No
additional FOT was detected for either of the patterned regions. We
emphasize that at the points in the phase diagram where the FOT
meets the matching fields, a contradictory behavior of the vortex
lattice occurs. On one hand, at the FOT there is a \emph{jump in
vortex density}. On the other hand, at matching fields the vortex
matter exhibits a \emph{strongly enhanced compressibility modulus}
$c_{11}\sim (dB/dH)^{-1}$ that exists both below and above $T_m$.
This apparent contradiction is discussed below.

\section{Discussion}
\begin{figure}[tb]
\includegraphics[clip,viewport=173 175 394 601,width=0.27\textwidth]{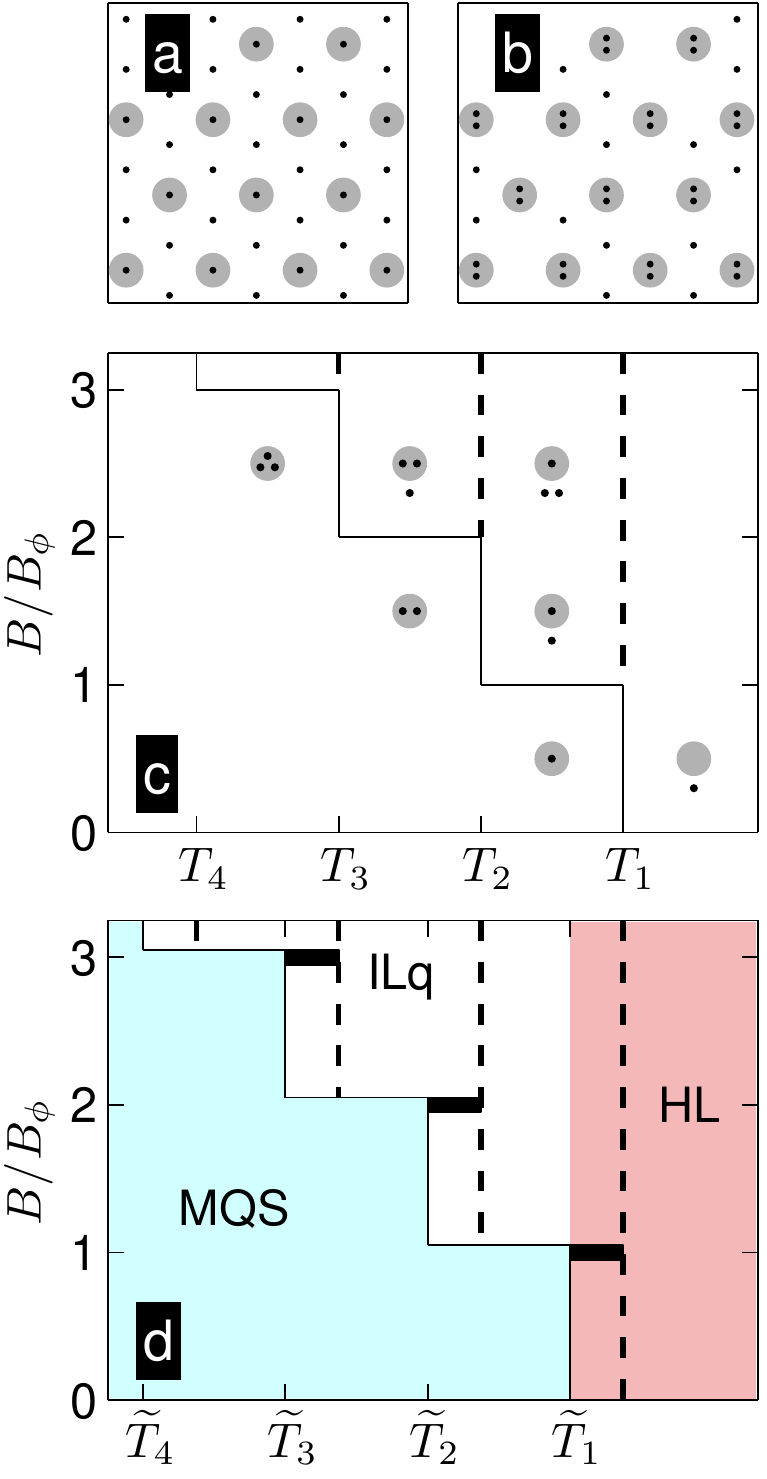}
\caption{(Color online) (a) and (b): A schematic view of the vortex
matter for $B/B_{\phi}=3$, within (a) the single-vortex pinning
scenario, and (b) the multi-quanta scenario. Gray circles indicate
holes. Black dots indicate vortices. (a) In each unit cell of the
pinning lattice, a single vortex is pinned to the hole and two
vortices are interstitial. Below (above) the IL interstitials are
pinned (mobile). (b) In each unit cell, two vortices remain pinned
to the hole. Below the IL the third vortex is also pinned to the hole
(not shown). Above the IL, it is depinned and mobile. (c) and (d): A
schematic description of the steps in the IL, within the
multi-quanta scenario, in the absence (c) and presence (d) of
thermal fluctuations or shaking. (c) The maximum number of quanta
per hole $n_{max}(T)$ (solid black line) is expected to decrease as
a function of $T$, resulting in temperatures $T_{n}$ (dashed lines)
above which holes pin $n-1$ vortices only, and additional vortices
become mobile interstitials. Thus for each interval
$(n-1)<B/B_{\phi}<n$, $T_{IL}=T_{n}$. (d) Thermal fluctuations shift
$T_{IL}$ (solid line) from $T_{n}$ (dashed lines) to
$\widetilde{T}_{n}<T_{n}$. At $B/B_{\phi}=n$, the vortex lattice
remains stable to the temperature $T_{n}$ due to a thermodynamic MI
phase. Resulting phases include a multi-quanta-solid (MSQ) below
$T_{IL}$, an interstitial liquid (ILq, white) above the IL that
terminates at $\widetilde{T}_{1}$, a homogeneous liquid (HL) above
$\widetilde{T}_{1}$, and MI ``fingers" at $B/B_{\phi}=n$
(black).}\label{fig:scenarios}
\end{figure}

In order to understand the observed behavior of the IL, we consider two possible
physical scenarios~\cite{prb:sto02}. In the first scenario, shown schematically in
Fig. \ref{fig:scenarios}(a), we assume that each hole can pin only a single vortex.
As a result, two vortex populations are present for $B>B_{\phi}$: vortices located
at holes, and interstitial vortices located between holes. The interstitial
vortices are subject to a caging potential caused by the vortices located at
holes~\cite{physc:kha93}, that is assumed to be weaker than the pinning potential
at holes, but stronger than the pristine pinning. This gives rise to a depinning
transition of the interstitials, which we identify with $T_{IL}$, at a temperature
above the pristine melting temperature $T_m$. Alternatively, we consider a scenario
in which there is multi-quanta pinning by holes. We assume that below the IL, all
vortices are located at holes, while above the IL, some vortices are depinned from
holes, and thus mobile, as shown schematically in Fig.~\ref{fig:scenarios}(b). Due
to repulsion between pinned vortices, the pinning force per vortex is expected to
decrease as a function of the number of vortices pinned to the
hole~\cite{jetp:mkr72}. We therefore assume that the pinned vortices residing at
holes depin one at a time, as $T$ is increased. Within this multi-quanta scenario,
$T_{IL}$ corresponds to the temperature at which the first vortices depin from
holes. Either of the two scenarios must provide an explanation for the observed
behavior of the IL: the plateaus in the IL \emph{away from} matching, the shift in
the IL to lower $T$ and $H$ in the presence of shaking, and the sharp dips
\emph{at} the matching fields.

We begin with the plateaus and steps in the IL. In the multi-quanta scenario, the
maximum number of vortices per hole $n_{max}(T)$ is determined by
temperature-dependent hole pinning strength and repulsive interactions between
vortices located at the hole, and decreases with increasing
temperature~\cite{jetp:mkr72}. We denote the temperature at which $n_{max}$
decreases from $n$ to $n-1$ by $T_{n}$ (see dotted lines in
Fig.~\ref{fig:scenarios}(c)). We assume that for the applied fields
$H\leq6B_{\phi}$, all vortices are pinned to holes at sufficiently low $T$. As $T$
is increased above $T_{n}$, holes may only pin $n-1$ vortices. Therefore vortices
abruptly depin from holes occupied by $n$ vortices, leaving $n-1$ vortices per
hole. The depinned vortices are mobile, leading to a fast onset of reversibility.
We therefore identify $T_{IL}=T_{n}$ for $(n-1)<B/B_{\phi}<n$. The resulting IL
thus displays steps and plateaus, as shown schematically in
Fig.~\ref{fig:scenarios}(c). Finite temperature and slight hole variability are
likely to cause some variation in $n_{max}(T)$ at different holes. This would lead
to some smearing of the irreversibility transition and to a weak dependence of
$T_{IL}$ on $B$ between matching fields due to different mixing of the $T_n$ as
field is varied. This schematic description neglects the effects of thermal
fluctuations, which will be discussed later on. The IL shown in
Fig.~\ref{fig:PD}(c) (white contour) indicates that $T_{IL}$ displays three
discrete steps in the temperature range $T=72-77~$K. The weak temperature
dependence of the IL between matching fields, as well as the plateaus in $dB/dH$ in
Figs.~\ref{fig:shakingEffect} and \ref{fig:higherOrders}, seem to indicate that the
IL is not strongly affected by interactions between vortices at neighboring holes.
Rather, it is governed by the hole pinning energy and the repulsion between
vortices within a single hole, leading to the step-like $T_{IL}$. Within the
single-vortex pinning scenario, in contrast, $T_{IL}(H)$ is the depinning
temperature of interstitial vortices, that is expected to decrease rather smoothly
with field as the density of interstitials increases.

We now address the the shift of the IL to lower temperatures in the
presence of shaking. Shaking and increased thermal fluctuations seem
to have a similar effect on the IL (see
Fig.~\ref{fig:higherOrders}). Within the multi-quanta scenario both
provide a mechanism for hopping of vortices between vacancies at
holes, thus increasing the dynamic $dB/dH$ and decreasing $T_{IL}$.
Between matching fields and slightly below $T_{IL}$, the holes are
below their full pinning capacity, so shaking may provide a way for
vortices to hop between holes more easily, or to depin from holes
and move to interstitial positions. The plateau-like behavior
indicates that this shaking-induced hopping is roughly independent
of the number of vacancies that exists. Instead, the degree of
hopping or depinning is dependent on the balance between the roughly
field-independent pinning energy of the $n_{max}$th vortex and the
activation energy of the applied shaking. Within the single-vortex
pinning scenario, above the IL the interstitials are mobile,
therefore immediately below the IL they are weakly pinned. Shaking
will thus assist in overcoming the weak pinning potential, decreasing
$T_{IL}$. The existence of the plateaus, however, cannot be easily
explained in this case.

We now address the third experimental finding, namely the
finger-like dips in $dB/dH$ \emph{at} matching fields. Focusing on
Figs.~\ref{fig:meltMatch} and \ref{fig:meltMatch2}, we see that
unlike the rest of the IL, the temperatures at which the fingers
terminate are \emph{not} affected by in-plane shaking. This strongly
suggests that unlike the IL, which is a \emph{dynamic} feature of
the phase diagram, the fingers at matching fields are a
\emph{thermodynamic} feature. A thermodynamic minimum with $dB/dH=0$
would indicate a plateau in the equilibrium $B(H)$ and a diverging
bulk modulus $c_{11}\sim
\partial{H}/\partial{B}$~~\cite{prb:wen98}, clear signatures of the
incompressible MI phase. The finite, positive minima observed in
$dB/dH$ (see Figs.~\ref{fig:shakingEffect}(b),
\ref{fig:shakingEffect}(c), and \ref{fig:higherOrders}) correspond
to a reduction in the positive slope of $B(H)$ at $nB_{\phi}$. These
finite values at $nB_{\phi}$ may be a consequence of the finite size
of the patterned region, which prevents infinite divergence of
$c_{11}$, or of the broadening of $dB/dH$ due to the modulation of
$\Delta H=1~$Oe, which is at least as wide as the width of the dip.
For samples with random defects, it has been argued that the MI
phase is destroyed by repulsive vortex interactions, possibly
retaining ``lock-in'' effects such as a finite peak in the bulk
modulus $c_{11}$~~\cite{prl:wen97,prb:wen98}. In our measurements,
however, the pinning is ordered. At matching fields, pinning energy
and vortex-vortex interactions both stabilize the vortex lattice. In
this case, observation of a MI phase is
possible~\cite{prl:wen97,prb:wen98}. We believe that the sharp
matching features we observed in $dB/dH$ at $nB_{\phi}$, that are
not affected by shaking, are a strong indication of thermodynamic MI
phases.

The fingers of reduced $dB/dH$, or MI phases, may be understood in
the context of the multi-quanta scenario. For $(n-1)<B/B_{\phi}<n$
and $T<T_{n}$, in the absence of thermal fluctuations, all vortices
are pinned to holes (see Fig.~\ref{fig:scenarios}(c)). The maximum
number of vortices that can be pinned $nB_{\phi}/\phi_0$ is greater
than the actual number of vortices $B/\phi_0$, resulting in
below-full occupancy, or ``vacancies", at some of the multi-quanta
holes. Thermal fluctuations (which were neglected in the schematic
description in Fig.~\ref{fig:scenarios}(c)) are expected to lead to
vortex hopping between the vacancies, resulting in enhanced vortex
mobility. Thus, for $(n-1)<B/B_{\phi}<n$, vortex dynamics lead to a
reduction in $T_{IL}$ from $T_{n}$ (Fig.~\ref{fig:scenarios}(d),
dashed lines) to some lower temperature $\widetilde{T}_{n}$. Exactly
at matching, an additional thermodynamic consideration enters. The
total hole capacity equals the number of vortices, and therefore
there is a finite energy cost for adding an extra interstitial
vortex. As a result, the equilibrium $B(H)$ will exhibit a plateau
over a finite range of $H$, that is not affected by shaking. The
corresponding minima in $dB/dH$, or equilibrium MI phases, may thus
be observed up to $T_{n}$ (black fingers in
Fig.~\ref{fig:scenarios}(d)).

The fingers of reduced $dB/dH$ may also be partially understood
within the single-vortex pinning scenario. At matching fields, we
expect both the vortices at holes and the interstitials to be ordered in
a configuration commensurate with the hole array, that enhances the
pinning potential and may result in a thermodynamic MI phase. Although
this picture is correct for commensurate matching fields
($B/B_{\phi}=1,3,4,7,$~etc. for a triangular array of
holes~\cite{prb:rei98}), it is not strictly correct for
incommensurate matching fields. At $B/B_{\phi}=2$, for example,
there are two interstitial positions with equal energy per unit
cell. This would imply that hopping of interstitials is possible
also \emph{at} incommensurate matching fields, resulting in greater
interstitial mobility, and hence a weaker matching effect. This is
\emph{not} what we observe in Fig.~\ref{fig:PD}(c), where the
fingers appear similar at the first and second matching fields.

We now consider the theoretical plausibility of the multi-quanta
scenario. The condition given for multiple-quanta pinning in holes
is~\cite{prb:buz93} $r^3>\xi \lambda^2$ for $\lambda \ll d$, where
$r$ is the hole radius, $d$ is the inter-hole distance, and
$\lambda=\lambda_0/\sqrt{1-T/T_c}$ is the London penetration
depth~\cite{revmp:bla94}. For $r=0.2~\mu$m, and
$\lambda_0=0.15~\mu$m~\cite{prl:jac95}, we may expect
multiple-quanta vortices for $T\leq88~$K for our sample.

We now estimate the number of vortices pinned at a hole as a
function of temperature. We consider two physical possibilities,
following Ref.~\cite{jetp:mkr72}. The first is that the $n$th vortex
is pinned as long as there is some force near the edge of the hole
that pushes it inwards. This corresponds to a pinned state that may
be metastable, depending on its free energy. The saturation number
\begin{equation}\label{eq:ns}
n_s\approx r/(2\xi)
\end{equation}
is the maximum $n$ for which this occurs. The second possibility for
multiple-quanta pinning is to require an equilibrium pinned state,
namely, the free energy of the $n$th vortex located in the hole must
be lower than its free energy far from the hole. In this case, the
number of pinned vortices is given by~\cite{jetp:mkr72}
\begin{equation}\label{eq:n0}
n_0 \approx \frac{1}{2} \ln{\frac{r}{2 \xi}} / \ln{\frac{2\lambda}{1.78 r}}.
\end{equation}
Both $n_s$ and $n_0$ decrease with $T$, which is consistent with the
observed downward steps in $H_{IL}(T)$. Substituting $r=0.2~\mu$m,
$\xi_0=2~$nm, $\lambda_0=0.15~\mu$m, and $T_c=90.5~$K, we obtain
$n_s=1-21$ for $T=90.46~$K down to $T=74.5~$K and $n_0=1-2$ for
$T=85.2~$K down to $T=76.1~$K. Note that Eq.~\ref{eq:n0} was derived
from the pinning energy of a \emph{single} hole. In the case of an
\emph{array} of holes, one should compare the free energy of a
pinned $n$th vortex to its free energy at the midpoint between two
neighboring holes. At this midpoint, there are positive
contributions to the free energy from the neighboring occupied
holes. Thus a higher free energy of the pinned vortex, or $n>n_0$,
would still be an equilibrium state of the system. Eq.~\ref{eq:n0}
should therefore be considered a lower limit on the equilibrium
number of vortices pinned to a hole. The fact that interactions
should raise estimated occupation numbers was also noted
by~\cite{prb:ber06}.

We now compare these theoretical estimates to the experimental
values of $n_{max}$. We observed three decreasing steps in $T_{IL}$
(see Fig.~\ref{fig:PD}(c)), with $\widetilde{T}_{6}\simeq72$~K,
$\widetilde{T}_{5}\simeq75$~K, and $\widetilde{T}_{4}\simeq77$~K.
According to the schematic phase diagram plotted in
Fig.~\ref{fig:scenarios}(d), the values of $\widetilde{T}_{n}$ are
lower than the temperatures $T_{n}$ at which the $n$-th multi-quanta
vortex depins in the absence of thermal fluctuations. From the
schematic description in Fig.~\ref{fig:scenarios}(c), we see that
the difference $T_{n}-\widetilde{T}_{n}$ may be estimated from the
difference in the $T_{IL}$ at, and slightly away from,
$B/B_{\phi}=n$. From Figs.~\ref{fig:meltMatch}(c) and
\ref{fig:meltMatch2}(c), we estimate this difference to be
$T_{n}-\widetilde{T}_{n}\sim2~$K. $T_6, T_5,$ and $T_4$ are thus
$74, 77$ and $79~$K, respectively, or equivalently,
$n_{max}(T<74)=6$, $n_{max}(74<T<77)=5$, $n_{max}(77<T<79)=4$, and
$n_{max}(T>79)=3$. We find that the extracted $n_{max}$ values are
much lower than the estimated $n_s$, and slightly higher than the
estimated $n_0$. Thus the equilibrium multi-quanta pinning scenario
described by Eq.~\ref{eq:n0} is more plausible. Note, however, that
Eqs.~\ref{eq:ns} and \ref{eq:n0} are based on the assumption of
fully penetrating holes; the number of vortices trapped by
surface holes may be lower~\cite{prb:bez96,prb:rae04}.

Summarizing our discussion of the IL, the multi-quanta scenario
provides a more consistent explanation as compared to single-vortex
pinning, both for the steps in the IL and for the similar MI fingers
at both commensurate and incommensurate matching fields. Indeed,
recent simulations of BSCCO with surface holes similar to those in
the experiment~\cite{mis:goldschmidt} indicate that multiple-quanta
occupation of surface holes does occur, and that depinning may occur
directly from holes.


Finally, we address the apparent contradiction of observing a FOT
within the patterned regions \emph{at matching fields}
$B/B_{\phi}=1,2$, as shown in Figs.~\ref{fig:meltMatch} and
\ref{fig:meltMatch2}. The observed FOT occurs at the same field and
temperature values as the pristine solid-liquid transition $T_m$,
and we therefore assume that both transitions are of similar nature.
This, however, seems to contradict the existence of MI phases at
matching fields, since in the MI phase the vortex lattice is
ordered, pinned, and incompressible, up to temperatures well above
the pristine $T_m$.

The observed FOT can be understood qualitatively by taking the bulk
beneath the surface holes into account. The surface holes are only
$\sim1.4~\mu$m deep, whereas the sample is $30~\mu$m thick. Although
the depth at which vortices are still sensitive to surface
patterning is not known exactly, magnetic decorations of BSCCO
crystals with square Fe periodic surface patterns indicate that the
hexagonal structure of the lattice is recovered fully just
$4.5~\mu$m beneath the pinning potential at low
temperatures~\cite{physc:fas04}. Thus one may expect that
sufficiently deep below the upper surface, the vortex ``tails" will
undergo a first-order melting transition at the pristine $T_m$. At
matching fields, however, the ``tips" of the vortices at the surface
are pinned to the periodic surface holes, and therefore behave as an
incompressible solid. The resulting situation immediately above
$T_m$ is rather unique: the vortex ``tips" are in a solid MI state,
while the vortex ``tails" are liquid. At the FOT, the vortex density
in the bulk increases by $\Delta B/\phi_0$, where $\Delta B$ is the
typical step in $B$ at $T_m$. If the ``tips" of the vortices were
fully incompressible, this $\Delta B$ in the bulk would be
completely shielded and unobservable at the top surface. Our data
indicate, however, that within our experimental range of parameters
the compressibility is high, but finite. Hence the MI top layer is
sufficiently transparent to allow observation of the paramagnetic
peak at the FOT of the underlying vortex ``tails". We conclude that
upon increasing temperature, the FOT observed within the patterned
region at matching fields indicates a transition from a solid bulk
with an incompressible solid surface to a unique state of an
essentially incompressible solid crust of vortex ``tips" concealing
a vortex liquid in the bulk.

It is interesting to note that this solid crust could be used as a
tool to study the interesting possibility of surface melting that
was suggested to exist in the vortex
lattice~\cite{nat:soi00,prl:dec06}, similar to surface wetting in
atomic solids.
This will require investigating crystals of different thicknesses
and samples patterned with holes at both the top and bottom
surfaces.

No additional FOTs were observed, even in the presence of shaking.
This indicates that the depinning line of the interstitial vortices,
and the delocalization line of the vortices pinned to holes, are
apparently \emph{not} FOTs. More accurate magnetization measurements
in the presence of shaking are needed to check for the existence of
second-order thermodynamic transitions.

\section{Conclusions}
BSCCO crystals with arrays of surface holes were measured using
differential magneto-optics accompanied by an in-plane shaking
field. We observe reduced $dB/dH$ of the patterned regions at
integer matching fields $B=nB_{\phi}$. These features are extremely
narrow, with a width in $H$ of less than $1~$Oe. The region of
reduced $dB/dH$ extends up to $87~\mathrm{K}$ for the first matching
field, terminating $3.5~\mathrm{K}$ below $T_c$. Shaking allowed the
equilibrium matching feature to be observed both above and below the
pristine melting line of the sample. This observation is in contrast
to previous dynamic measurements of BSCCO crystals with periodic
surface holes~\cite{physc:ooi05,physc:ooi06}, in which matching
effects were observed only well above the pristine melting line. Our
finding of a sharply suppressed compressibility of an equilibrated
vortex lattice at the matching fields is a strong evidence of the
existence of Mott insulator phases.

We observe a first-order melting transition within the patterned
areas, both away from matching and at the first and second matching
fields. Surprisingly, this transition is not shifted with respect to
the pristine melting transition. We interpret this transition as
first-order melting of the vortices in the pristine bulk beneath the
patterned surface, that results in a step in vortex density. The
added vortex ``tails" beneath the patterned surface force vortex
``tips" through the surface, even in regions of reduced
compressibility. We emphasize that this is an extremely unusual
situation, in which the vortex tails located in the bulk are in a
liquid phase, while their tips located near the surface are in a
pinned, ordered phase, and therefore solid.

The irreversibility line of the patterned regions is found to be
shifted upward in $H$ and $T$ to above the pristine melting line
$T_m$, and displays step-like behavior, with almost no temperature
dependence between matching fields $B=nB_{\phi}$. These steps are
consistent with a multi-vortex pinning scenario, with an estimated
maximum value of $n_{max}=6$ flux quanta pinned to each surface hole
at $H/B_{\phi}\gtrsim 5$ and $T\leq72$~K.

Applied shaking shifts the irreversibility line to lower field and temperature,
enabling the observation of the first-order transition. However, no first-order
transitions related to the surface holes, corresponding to depinning lines of
interstitial vortices, or delocalization of vortices pinned to holes, were
observed. Further transport measurements are necessary to determine if an
additional delocalization line exists at higher temperatures, in analogy to the
delocalization line of the vortices residing on columnar defects, that separates
the interstitial liquid from a homogeneous liquid~\cite{prl:ban04}.

\begin{acknowledgments}

We wish to thank A. Lahav for technical assistance. We are grateful to Y.
Goldschmidt and M. Opferman for providing insights from simulations. We thank
V. B. Geshkenbein, E. H. Brandt, and G. P. Mikitik for helpful discussions.
This work was supported by the US-Israel Binational Science Foundation (BSF),
by the Minerva Foundation, and by Grant-in-aid from the Ministry of Education,
Culture, Sport, Science and Technology, Japan. E.Z. acknowledges the support of
the German Israeli Foundation (GIF).

\end{acknowledgments}


\begin{thebibliography}{64}
\expandafter\ifx\csname natexlab\endcsname\relax\def\natexlab#1{#1}\fi
\expandafter\ifx\csname bibnamefont\endcsname\relax
  \def\bibnamefont#1{#1}\fi
\expandafter\ifx\csname bibfnamefont\endcsname\relax
  \def\bibfnamefont#1{#1}\fi
\expandafter\ifx\csname citenamefont\endcsname\relax
  \def\citenamefont#1{#1}\fi
\expandafter\ifx\csname url\endcsname\relax
  \def\url#1{\texttt{#1}}\fi
\expandafter\ifx\csname urlprefix\endcsname\relax\def\urlprefix{URL }\fi
\providecommand{\bibinfo}[2]{#2}
\providecommand{\eprint}[2][]{\url{#2}}

\bibitem[{\citenamefont{Civale et~al.}(1991)\citenamefont{Civale, Marwick,
  Worthington, Kirk, Thompson, Krusin-Elbaum, Sun, Clem, and
  Holtzberg}}]{prl:civ91}
\bibinfo{author}{\bibfnamefont{L.}~\bibnamefont{Civale}},
  \bibinfo{author}{\bibfnamefont{A.~D.} \bibnamefont{Marwick}},
  \bibinfo{author}{\bibfnamefont{T.~K.} \bibnamefont{Worthington}},
  \bibinfo{author}{\bibfnamefont{M.~A.} \bibnamefont{Kirk}},
  \bibinfo{author}{\bibfnamefont{J.~R.} \bibnamefont{Thompson}},
  \bibinfo{author}{\bibfnamefont{L.}~\bibnamefont{Krusin-Elbaum}},
  \bibinfo{author}{\bibfnamefont{Y.}~\bibnamefont{Sun}},
  \bibinfo{author}{\bibfnamefont{J.~R.} \bibnamefont{Clem}}, \bibnamefont{and}
  \bibinfo{author}{\bibfnamefont{F.}~\bibnamefont{Holtzberg}},
  \bibinfo{journal}{Phys. Rev. Lett.} \textbf{\bibinfo{volume}{67}},
  \bibinfo{pages}{648} (\bibinfo{year}{1991}).

\bibitem[{\citenamefont{Konczykowski et~al.}(1991)\citenamefont{Konczykowski,
  Rullier-Albenque, Yacoby, Shaulov, Yeshurun, and Lejay}}]{prb:mar91}
\bibinfo{author}{\bibfnamefont{M.}~\bibnamefont{Konczykowski}},
  \bibinfo{author}{\bibfnamefont{F.}~\bibnamefont{Rullier-Albenque}},
  \bibinfo{author}{\bibfnamefont{E.~R.} \bibnamefont{Yacoby}},
  \bibinfo{author}{\bibfnamefont{A.}~\bibnamefont{Shaulov}},
  \bibinfo{author}{\bibfnamefont{Y.}~\bibnamefont{Yeshurun}}, \bibnamefont{and}
  \bibinfo{author}{\bibfnamefont{P.}~\bibnamefont{Lejay}},
  \bibinfo{journal}{Phys. Rev. B} \textbf{\bibinfo{volume}{44}},
  \bibinfo{pages}{7167} (\bibinfo{year}{1991}).

\bibitem[{\citenamefont{Nelson and Vinokur}(1993)}]{prb:nel93}
\bibinfo{author}{\bibfnamefont{D.~R.} \bibnamefont{Nelson}} \bibnamefont{and}
  \bibinfo{author}{\bibfnamefont{V.~M.} \bibnamefont{Vinokur}},
  \bibinfo{journal}{Phys. Rev. B} \textbf{\bibinfo{volume}{48}},
  \bibinfo{pages}{13060} (\bibinfo{year}{1993}).

\bibitem[{\citenamefont{Blatter et~al.}(1994)\citenamefont{Blatter, Feigel'man,
  Geshkenbein, Larkin, and Vinokur}}]{revmp:bla94}
\bibinfo{author}{\bibfnamefont{G.}~\bibnamefont{Blatter}},
  \bibinfo{author}{\bibfnamefont{M.~V.} \bibnamefont{Feigel'man}},
  \bibinfo{author}{\bibfnamefont{V.~B.} \bibnamefont{Geshkenbein}},
  \bibinfo{author}{\bibfnamefont{A.~I.} \bibnamefont{Larkin}},
  \bibnamefont{and} \bibinfo{author}{\bibfnamefont{V.~M.}
  \bibnamefont{Vinokur}}, \bibinfo{journal}{Rev. Mod. Phys.}
  \textbf{\bibinfo{volume}{66}}, \bibinfo{pages}{1125} (\bibinfo{year}{1994}).

\bibitem[{\citenamefont{Radzihovsky}(1995)}]{prl:rad95}
\bibinfo{author}{\bibfnamefont{L.}~\bibnamefont{Radzihovsky}},
  \bibinfo{journal}{Phys. Rev. Lett.} \textbf{\bibinfo{volume}{74}},
  \bibinfo{pages}{4923} (\bibinfo{year}{1995}).

\bibitem[{\citenamefont{Lopatin and Vinokur}(2004)}]{prl:lop04}
\bibinfo{author}{\bibfnamefont{A.~V.} \bibnamefont{Lopatin}} \bibnamefont{and}
  \bibinfo{author}{\bibfnamefont{V.~M.} \bibnamefont{Vinokur}},
  \bibinfo{journal}{Phys. Rev. Lett.} \textbf{\bibinfo{volume}{92}},
  \bibinfo{pages}{067008} (\bibinfo{year}{2004}).

\bibitem[{\citenamefont{Tyagi and Goldschmidt}(2003)}]{prb:tya03}
\bibinfo{author}{\bibfnamefont{S.}~\bibnamefont{Tyagi}} \bibnamefont{and}
  \bibinfo{author}{\bibfnamefont{Y.~Y.} \bibnamefont{Goldschmidt}},
  \bibinfo{journal}{Phys. Rev. B} \textbf{\bibinfo{volume}{67}},
  \bibinfo{pages}{214501} (\bibinfo{year}{2003}).

\bibitem[{\citenamefont{Dasgupta and Valls}(2003)}]{prl:das03}
\bibinfo{author}{\bibfnamefont{C.}~\bibnamefont{Dasgupta}} \bibnamefont{and}
  \bibinfo{author}{\bibfnamefont{O.~T.} \bibnamefont{Valls}},
  \bibinfo{journal}{Phys. Rev. Lett.} \textbf{\bibinfo{volume}{91}},
  \bibinfo{pages}{127002} (\bibinfo{year}{2003}).

\bibitem[{\citenamefont{van~der Beek et~al.}(2001)\citenamefont{van~der Beek,
  Konczykowski, Samoilov, Chikumoto, Bouffard, and Feigel'man}}]{prl:van01}
\bibinfo{author}{\bibfnamefont{C.~J.} \bibnamefont{van~der Beek}},
  \bibinfo{author}{\bibfnamefont{M.}~\bibnamefont{Konczykowski}},
  \bibinfo{author}{\bibfnamefont{A.~V.} \bibnamefont{Samoilov}},
  \bibinfo{author}{\bibfnamefont{N.}~\bibnamefont{Chikumoto}},
  \bibinfo{author}{\bibfnamefont{S.}~\bibnamefont{Bouffard}}, \bibnamefont{and}
  \bibinfo{author}{\bibfnamefont{M.~V.} \bibnamefont{Feigel'man}},
  \bibinfo{journal}{Phys. Rev. Lett.} \textbf{\bibinfo{volume}{86}},
  \bibinfo{pages}{5136} (\bibinfo{year}{2001}).

\bibitem[{\citenamefont{Banerjee et~al.}(2003)\citenamefont{Banerjee, Soibel,
  Myasoedov, Rappaport, Zeldov, Menghini, Fasano, de~la Cruz, van~der Beek,
  Konczykowski et~al.}}]{prl:ban03}
\bibinfo{author}{\bibfnamefont{S.~S.} \bibnamefont{Banerjee}},
  \bibinfo{author}{\bibfnamefont{A.}~\bibnamefont{Soibel}},
  \bibinfo{author}{\bibfnamefont{Y.}~\bibnamefont{Myasoedov}},
  \bibinfo{author}{\bibfnamefont{M.}~\bibnamefont{Rappaport}},
  \bibinfo{author}{\bibfnamefont{E.}~\bibnamefont{Zeldov}},
  \bibinfo{author}{\bibfnamefont{M.}~\bibnamefont{Menghini}},
  \bibinfo{author}{\bibfnamefont{Y.}~\bibnamefont{Fasano}},
  \bibinfo{author}{\bibfnamefont{F.}~\bibnamefont{de~la Cruz}},
  \bibinfo{author}{\bibfnamefont{C.~J.} \bibnamefont{van~der Beek}},
  \bibinfo{author}{\bibfnamefont{M.}~\bibnamefont{Konczykowski}},
  \bibnamefont{et~al.}, \bibinfo{journal}{Phys. Rev. Lett.}
  \textbf{\bibinfo{volume}{90}}, \bibinfo{pages}{087004}
  (\bibinfo{year}{2003}).

\bibitem[{\citenamefont{Menghini et~al.}(2003)\citenamefont{Menghini, Fasano,
  de~la Cruz, Banerjee, Myasoedov, Zeldov, van~der Beek, Konczykowski, and
  Tamegai}}]{prl:men03}
\bibinfo{author}{\bibfnamefont{M.}~\bibnamefont{Menghini}},
  \bibinfo{author}{\bibfnamefont{Y.}~\bibnamefont{Fasano}},
  \bibinfo{author}{\bibfnamefont{F.}~\bibnamefont{de~la Cruz}},
  \bibinfo{author}{\bibfnamefont{S.~S.} \bibnamefont{Banerjee}},
  \bibinfo{author}{\bibfnamefont{Y.}~\bibnamefont{Myasoedov}},
  \bibinfo{author}{\bibfnamefont{E.}~\bibnamefont{Zeldov}},
  \bibinfo{author}{\bibfnamefont{C.~J.} \bibnamefont{van~der Beek}},
  \bibinfo{author}{\bibfnamefont{M.}~\bibnamefont{Konczykowski}},
  \bibnamefont{and} \bibinfo{author}{\bibfnamefont{T.}~\bibnamefont{Tamegai}},
  \bibinfo{journal}{Phys. Rev. Lett.} \textbf{\bibinfo{volume}{90}},
  \bibinfo{pages}{147001} (\bibinfo{year}{2003}).

\bibitem[{\citenamefont{Banerjee et~al.}(2004)\citenamefont{Banerjee, Goldberg,
  Soibel, Myasoedov, Rappaport, Zeldov, de~la Cruz, van~der Beek, Konczykowski,
  Tamegai et~al.}}]{prl:ban04}
\bibinfo{author}{\bibfnamefont{S.~S.} \bibnamefont{Banerjee}},
  \bibinfo{author}{\bibfnamefont{S.}~\bibnamefont{Goldberg}},
  \bibinfo{author}{\bibfnamefont{A.}~\bibnamefont{Soibel}},
  \bibinfo{author}{\bibfnamefont{Y.}~\bibnamefont{Myasoedov}},
  \bibinfo{author}{\bibfnamefont{M.}~\bibnamefont{Rappaport}},
  \bibinfo{author}{\bibfnamefont{E.}~\bibnamefont{Zeldov}},
  \bibinfo{author}{\bibfnamefont{F.}~\bibnamefont{de~la Cruz}},
  \bibinfo{author}{\bibfnamefont{C.~J.} \bibnamefont{van~der Beek}},
  \bibinfo{author}{\bibfnamefont{M.}~\bibnamefont{Konczykowski}},
  \bibinfo{author}{\bibfnamefont{T.}~\bibnamefont{Tamegai}},
  \bibnamefont{et~al.}, \bibinfo{journal}{Phys. Rev. Lett.}
  \textbf{\bibinfo{volume}{93}}, \bibinfo{pages}{097002}
  (\bibinfo{year}{2004}).

\bibitem[{\citenamefont{Fiory et~al.}(1978)\citenamefont{Fiory, Hebard, and
  Somekh}}]{apl:fio73}
\bibinfo{author}{\bibfnamefont{A.~T.} \bibnamefont{Fiory}},
  \bibinfo{author}{\bibfnamefont{A.~F.} \bibnamefont{Hebard}},
  \bibnamefont{and} \bibinfo{author}{\bibfnamefont{S.}~\bibnamefont{Somekh}},
  \bibinfo{journal}{Appl. Phys. Lett.} \textbf{\bibinfo{volume}{32}},
  \bibinfo{pages}{73} (\bibinfo{year}{1978}).

\bibitem[{\citenamefont{Lykov}(1993)}]{ssc:lyk93}
\bibinfo{author}{\bibfnamefont{A.~N.} \bibnamefont{Lykov}},
  \bibinfo{journal}{Solid State Commun.} \textbf{\bibinfo{volume}{86}},
  \bibinfo{pages}{531} (\bibinfo{year}{1993}).

\bibitem[{\citenamefont{Morgan and Ketterson}(1998)}]{prl:mor98}
\bibinfo{author}{\bibfnamefont{D.~J.} \bibnamefont{Morgan}} \bibnamefont{and}
  \bibinfo{author}{\bibfnamefont{J.~B.} \bibnamefont{Ketterson}},
  \bibinfo{journal}{Phys. Rev. Lett.} \textbf{\bibinfo{volume}{80}},
  \bibinfo{pages}{3614} (\bibinfo{year}{1998}).

\bibitem[{\citenamefont{Moshchalkov et~al.}(1998)\citenamefont{Moshchalkov,
  Baert, Metlushko, Rosseel, Van~Bael, Temst, Bruynseraede, and
  Jonckheere}}]{prb:mos98}
\bibinfo{author}{\bibfnamefont{V.~V.} \bibnamefont{Moshchalkov}},
  \bibinfo{author}{\bibfnamefont{M.}~\bibnamefont{Baert}},
  \bibinfo{author}{\bibfnamefont{V.~V.} \bibnamefont{Metlushko}},
  \bibinfo{author}{\bibfnamefont{E.}~\bibnamefont{Rosseel}},
  \bibinfo{author}{\bibfnamefont{M.~J.} \bibnamefont{Van~Bael}},
  \bibinfo{author}{\bibfnamefont{K.}~\bibnamefont{Temst}},
  \bibinfo{author}{\bibfnamefont{Y.}~\bibnamefont{Bruynseraede}},
  \bibnamefont{and}
  \bibinfo{author}{\bibfnamefont{R.}~\bibnamefont{Jonckheere}},
  \bibinfo{journal}{Phys. Rev. B} \textbf{\bibinfo{volume}{57}},
  \bibinfo{pages}{3615} (\bibinfo{year}{1998}).

\bibitem[{\citenamefont{Baert et~al.}(1995)\citenamefont{Baert, Metlushko,
  Jonckheere, Moshchalkov, and Bruynseraede}}]{prl:bae95}
\bibinfo{author}{\bibfnamefont{M.}~\bibnamefont{Baert}},
  \bibinfo{author}{\bibfnamefont{V.~V.} \bibnamefont{Metlushko}},
  \bibinfo{author}{\bibfnamefont{R.}~\bibnamefont{Jonckheere}},
  \bibinfo{author}{\bibfnamefont{V.~V.} \bibnamefont{Moshchalkov}},
  \bibnamefont{and}
  \bibinfo{author}{\bibfnamefont{Y.}~\bibnamefont{Bruynseraede}},
  \bibinfo{journal}{Phys. Rev. Lett.} \textbf{\bibinfo{volume}{74}},
  \bibinfo{pages}{3269} (\bibinfo{year}{1995}).

\bibitem[{\citenamefont{Mart\'in et~al.}(1997)\citenamefont{Mart\'in, V\'elez,
  Nogu\'es, and Schuller}}]{prl:mar97}
\bibinfo{author}{\bibfnamefont{J.~I.} \bibnamefont{Mart\'in}},
  \bibinfo{author}{\bibfnamefont{M.}~\bibnamefont{V\'elez}},
  \bibinfo{author}{\bibfnamefont{J.}~\bibnamefont{Nogu\'es}}, \bibnamefont{and}
  \bibinfo{author}{\bibfnamefont{I.~K.} \bibnamefont{Schuller}},
  \bibinfo{journal}{Phys. Rev. Lett.} \textbf{\bibinfo{volume}{79}},
  \bibinfo{pages}{1929} (\bibinfo{year}{1997}).

\bibitem[{\citenamefont{Rosseel et~al.}(1996)\citenamefont{Rosseel, Van~Bael,
  Baert, Jonckheere, Moshchalkov, and Bruynseraede}}]{prb:ros96}
\bibinfo{author}{\bibfnamefont{E.}~\bibnamefont{Rosseel}},
  \bibinfo{author}{\bibfnamefont{M.}~\bibnamefont{Van~Bael}},
  \bibinfo{author}{\bibfnamefont{M.}~\bibnamefont{Baert}},
  \bibinfo{author}{\bibfnamefont{R.}~\bibnamefont{Jonckheere}},
  \bibinfo{author}{\bibfnamefont{V.~V.} \bibnamefont{Moshchalkov}},
  \bibnamefont{and}
  \bibinfo{author}{\bibfnamefont{Y.}~\bibnamefont{Bruynseraede}},
  \bibinfo{journal}{Phys. Rev. B} \textbf{\bibinfo{volume}{53}},
  \bibinfo{pages}{R2983} (\bibinfo{year}{1996}).

\bibitem[{\citenamefont{Metlushko et~al.}(1998)\citenamefont{Metlushko, {De
  Long}, Baert, Rosseel, Van~Bael, Temst, Moshchalkov, and
  Bruynseraede}}]{epl:met98}
\bibinfo{author}{\bibfnamefont{V.~V.} \bibnamefont{Metlushko}},
  \bibinfo{author}{\bibfnamefont{L.~E.} \bibnamefont{{De Long}}},
  \bibinfo{author}{\bibfnamefont{M.}~\bibnamefont{Baert}},
  \bibinfo{author}{\bibfnamefont{E.}~\bibnamefont{Rosseel}},
  \bibinfo{author}{\bibfnamefont{M.~J.} \bibnamefont{Van~Bael}},
  \bibinfo{author}{\bibfnamefont{K.}~\bibnamefont{Temst}},
  \bibinfo{author}{\bibfnamefont{V.~V.} \bibnamefont{Moshchalkov}},
  \bibnamefont{and}
  \bibinfo{author}{\bibfnamefont{Y.}~\bibnamefont{Bruynseraede}},
  \bibinfo{journal}{Europhys. Lett.} \textbf{\bibinfo{volume}{41}},
  \bibinfo{pages}{333} (\bibinfo{year}{1998}).

\bibitem[{\citenamefont{Metlushko et~al.}(1999)\citenamefont{Metlushko, Welp,
  Crabtree, Zhang, Brueck, Watkins, DeLong, Ilic, Chung, and
  Hesketh}}]{prb:met99}
\bibinfo{author}{\bibfnamefont{V.}~\bibnamefont{Metlushko}},
  \bibinfo{author}{\bibfnamefont{U.}~\bibnamefont{Welp}},
  \bibinfo{author}{\bibfnamefont{G.~W.} \bibnamefont{Crabtree}},
  \bibinfo{author}{\bibfnamefont{Z.}~\bibnamefont{Zhang}},
  \bibinfo{author}{\bibfnamefont{S.~R.~J.} \bibnamefont{Brueck}},
  \bibinfo{author}{\bibfnamefont{B.}~\bibnamefont{Watkins}},
  \bibinfo{author}{\bibfnamefont{L.~E.} \bibnamefont{DeLong}},
  \bibinfo{author}{\bibfnamefont{B.}~\bibnamefont{Ilic}},
  \bibinfo{author}{\bibfnamefont{K.}~\bibnamefont{Chung}}, \bibnamefont{and}
  \bibinfo{author}{\bibfnamefont{P.~J.} \bibnamefont{Hesketh}},
  \bibinfo{journal}{Phys. Rev. B} \textbf{\bibinfo{volume}{59}},
  \bibinfo{pages}{603} (\bibinfo{year}{1999}).

\bibitem[{\citenamefont{{De Long} et~al.}(2002)\citenamefont{{De Long},
  Metlushko, Kryukov, Yun, Lokhre, Moshchalkov, and
  Bruynseraede}}]{physc:del02}
\bibinfo{author}{\bibfnamefont{L.~E.} \bibnamefont{{De Long}}},
  \bibinfo{author}{\bibfnamefont{V.~V.} \bibnamefont{Metlushko}},
  \bibinfo{author}{\bibfnamefont{S.}~\bibnamefont{Kryukov}},
  \bibinfo{author}{\bibfnamefont{M.}~\bibnamefont{Yun}},
  \bibinfo{author}{\bibfnamefont{S.}~\bibnamefont{Lokhre}},
  \bibinfo{author}{\bibfnamefont{V.~V.} \bibnamefont{Moshchalkov}},
  \bibnamefont{and}
  \bibinfo{author}{\bibfnamefont{Y.}~\bibnamefont{Bruynseraede}},
  \bibinfo{journal}{Physica C} \textbf{\bibinfo{volume}{369}},
  \bibinfo{pages}{118} (\bibinfo{year}{2002}).

\bibitem[{\citenamefont{Silhanek et~al.}(2004)\citenamefont{Silhanek, Raedts,
  Van~Bael, and Moshchalkov}}]{prb:sil04}
\bibinfo{author}{\bibfnamefont{A.~V.} \bibnamefont{Silhanek}},
  \bibinfo{author}{\bibfnamefont{S.}~\bibnamefont{Raedts}},
  \bibinfo{author}{\bibfnamefont{M.~J.} \bibnamefont{Van~Bael}},
  \bibnamefont{and} \bibinfo{author}{\bibfnamefont{V.~V.}
  \bibnamefont{Moshchalkov}}, \bibinfo{journal}{Phys. Rev. B}
  \textbf{\bibinfo{volume}{70}}, \bibinfo{pages}{054515}
  (\bibinfo{year}{2004}).

\bibitem[{\citenamefont{Khalfin and Shapiro}(1995)}]{physc:kha93}
\bibinfo{author}{\bibfnamefont{I.~B.} \bibnamefont{Khalfin}} \bibnamefont{and}
  \bibinfo{author}{\bibfnamefont{B.~Y.} \bibnamefont{Shapiro}},
  \bibinfo{journal}{Physica C} \textbf{\bibinfo{volume}{207}},
  \bibinfo{pages}{359} (\bibinfo{year}{1995}).

\bibitem[{\citenamefont{Reichhardt et~al.}(1998)\citenamefont{Reichhardt,
  Olson, and Nori}}]{prb:rei98}
\bibinfo{author}{\bibfnamefont{C.}~\bibnamefont{Reichhardt}},
  \bibinfo{author}{\bibfnamefont{C.~J.} \bibnamefont{Olson}}, \bibnamefont{and}
  \bibinfo{author}{\bibfnamefont{F.}~\bibnamefont{Nori}},
  \bibinfo{journal}{Phys. Rev. B} \textbf{\bibinfo{volume}{57}},
  \bibinfo{pages}{7937} (\bibinfo{year}{1998}).

\bibitem[{\citenamefont{Doria and de~Andrade}(1999)}]{prb:dor99}
\bibinfo{author}{\bibfnamefont{M.~M.} \bibnamefont{Doria}} \bibnamefont{and}
  \bibinfo{author}{\bibfnamefont{S.~C.~B.} \bibnamefont{de~Andrade}},
  \bibinfo{journal}{Phys. Rev. B} \textbf{\bibinfo{volume}{60}},
  \bibinfo{pages}{13164} (\bibinfo{year}{1999}).

\bibitem[{\citenamefont{Berdiyorov et~al.}(2006)\citenamefont{Berdiyorov,
  Milo\v{s}evi\'{c}, and Peeters}}]{prb:ber06}
\bibinfo{author}{\bibfnamefont{G.~R.} \bibnamefont{Berdiyorov}},
  \bibinfo{author}{\bibfnamefont{M.~V.} \bibnamefont{Milo\v{s}evi\'{c}}},
  \bibnamefont{and} \bibinfo{author}{\bibfnamefont{F.~M.}
  \bibnamefont{Peeters}}, \bibinfo{journal}{Phys. Rev. B}
  \textbf{\bibinfo{volume}{74}}, \bibinfo{eid}{174512} (\bibinfo{year}{2006}).

\bibitem[{\citenamefont{Laguna et~al.}(2001)\citenamefont{Laguna, Balseiro,
  Dom\'inguez, and Nori}}]{prb:lag01}
\bibinfo{author}{\bibfnamefont{M.~F.} \bibnamefont{Laguna}},
  \bibinfo{author}{\bibfnamefont{C.~A.} \bibnamefont{Balseiro}},
  \bibinfo{author}{\bibfnamefont{D.}~\bibnamefont{Dom\'inguez}},
  \bibnamefont{and} \bibinfo{author}{\bibfnamefont{F.}~\bibnamefont{Nori}},
  \bibinfo{journal}{Phys. Rev. B} \textbf{\bibinfo{volume}{64}},
  \bibinfo{pages}{104505} (\bibinfo{year}{2001}).

\bibitem[{\citenamefont{Reichhardt et~al.}(2001)\citenamefont{Reichhardt,
  Olson, Scalettar, and Zim\'anyi}}]{prb:rei01}
\bibinfo{author}{\bibfnamefont{C.}~\bibnamefont{Reichhardt}},
  \bibinfo{author}{\bibfnamefont{C.~J.} \bibnamefont{Olson}},
  \bibinfo{author}{\bibfnamefont{R.~T.} \bibnamefont{Scalettar}},
  \bibnamefont{and} \bibinfo{author}{\bibfnamefont{G.~T.}
  \bibnamefont{Zim\'anyi}}, \bibinfo{journal}{Phys. Rev. B}
  \textbf{\bibinfo{volume}{64}}, \bibinfo{pages}{144509}
  (\bibinfo{year}{2001}).

\bibitem[{\citenamefont{Harada et~al.}(1996)\citenamefont{Harada, Kamimura,
  Kasai, Matsuda, Tonomura, and Moshchalkov}}]{sci:har96}
\bibinfo{author}{\bibfnamefont{K.}~\bibnamefont{Harada}},
  \bibinfo{author}{\bibfnamefont{O.}~\bibnamefont{Kamimura}},
  \bibinfo{author}{\bibfnamefont{H.}~\bibnamefont{Kasai}},
  \bibinfo{author}{\bibfnamefont{T.}~\bibnamefont{Matsuda}},
  \bibinfo{author}{\bibfnamefont{A.}~\bibnamefont{Tonomura}}, \bibnamefont{and}
  \bibinfo{author}{\bibfnamefont{V.~V.} \bibnamefont{Moshchalkov}},
  \bibinfo{journal}{Science} \textbf{\bibinfo{volume}{274}},
  \bibinfo{pages}{1167} (\bibinfo{year}{1996}).

\bibitem[{\citenamefont{Field et~al.}(2002)\citenamefont{Field, James,
  Barentine, Metlushko, Crabtree, Shtrikman, Ilic, and Brueck}}]{prl:fie02}
\bibinfo{author}{\bibfnamefont{S.~B.} \bibnamefont{Field}},
  \bibinfo{author}{\bibfnamefont{S.~S.} \bibnamefont{James}},
  \bibinfo{author}{\bibfnamefont{J.}~\bibnamefont{Barentine}},
  \bibinfo{author}{\bibfnamefont{V.}~\bibnamefont{Metlushko}},
  \bibinfo{author}{\bibfnamefont{G.}~\bibnamefont{Crabtree}},
  \bibinfo{author}{\bibfnamefont{H.}~\bibnamefont{Shtrikman}},
  \bibinfo{author}{\bibfnamefont{B.}~\bibnamefont{Ilic}}, \bibnamefont{and}
  \bibinfo{author}{\bibfnamefont{S.~R.~J.} \bibnamefont{Brueck}},
  \bibinfo{journal}{Phys. Rev. Lett.} \textbf{\bibinfo{volume}{88}},
  \bibinfo{pages}{067003} (\bibinfo{year}{2002}).

\bibitem[{\citenamefont{Van~Look et~al.}(1999)\citenamefont{Van~Look, Rosseel,
  Van~Bael, Temst, Moshchalkov, and Bruynseraede}}]{prb:van99}
\bibinfo{author}{\bibfnamefont{L.}~\bibnamefont{Van~Look}},
  \bibinfo{author}{\bibfnamefont{E.}~\bibnamefont{Rosseel}},
  \bibinfo{author}{\bibfnamefont{M.~J.} \bibnamefont{Van~Bael}},
  \bibinfo{author}{\bibfnamefont{K.}~\bibnamefont{Temst}},
  \bibinfo{author}{\bibfnamefont{V.~V.} \bibnamefont{Moshchalkov}},
  \bibnamefont{and}
  \bibinfo{author}{\bibfnamefont{Y.}~\bibnamefont{Bruynseraede}},
  \bibinfo{journal}{Phys. Rev. B} \textbf{\bibinfo{volume}{60}},
  \bibinfo{pages}{R6998} (\bibinfo{year}{1999}).

\bibitem[{\citenamefont{Castellanos et~al.}(1997)\citenamefont{Castellanos,
  W\"{o}rdenweber, Ockenfuss, v.d. Hart, and Keck}}]{apl:cas97}
\bibinfo{author}{\bibfnamefont{A.}~\bibnamefont{Castellanos}},
  \bibinfo{author}{\bibfnamefont{R.}~\bibnamefont{W\"{o}rdenweber}},
  \bibinfo{author}{\bibfnamefont{G.}~\bibnamefont{Ockenfuss}},
  \bibinfo{author}{\bibfnamefont{A.}~\bibnamefont{v.d. Hart}},
  \bibnamefont{and} \bibinfo{author}{\bibfnamefont{K.}~\bibnamefont{Keck}},
  \bibinfo{journal}{Appl. Phys. Lett.} \textbf{\bibinfo{volume}{71}},
  \bibinfo{pages}{962} (\bibinfo{year}{1997}).

\bibitem[{\citenamefont{Crisan et~al.}(2005)\citenamefont{Crisan, Pross, Cole,
  Bending, W\"{o}rdenweber, Lahl, and Brandt}}]{prb:cri05}
\bibinfo{author}{\bibfnamefont{A.}~\bibnamefont{Crisan}},
  \bibinfo{author}{\bibfnamefont{A.}~\bibnamefont{Pross}},
  \bibinfo{author}{\bibfnamefont{D.}~\bibnamefont{Cole}},
  \bibinfo{author}{\bibfnamefont{S.~J.} \bibnamefont{Bending}},
  \bibinfo{author}{\bibfnamefont{R.}~\bibnamefont{W\"{o}rdenweber}},
  \bibinfo{author}{\bibfnamefont{P.}~\bibnamefont{Lahl}}, \bibnamefont{and}
  \bibinfo{author}{\bibfnamefont{E.~H.} \bibnamefont{Brandt}},
  \bibinfo{journal}{Phys. Rev. B} \textbf{\bibinfo{volume}{71}},
  \bibinfo{eid}{144504} (\bibinfo{year}{2005}).

\bibitem[{\citenamefont{Ooi et~al.}(2006)\citenamefont{Ooi, Mochiku, Ishii, Yu,
  and Hirata}}]{physc:ooi06}
\bibinfo{author}{\bibfnamefont{S.}~\bibnamefont{Ooi}},
  \bibinfo{author}{\bibfnamefont{T.}~\bibnamefont{Mochiku}},
  \bibinfo{author}{\bibfnamefont{S.}~\bibnamefont{Ishii}},
  \bibinfo{author}{\bibfnamefont{S.}~\bibnamefont{Yu}}, \bibnamefont{and}
  \bibinfo{author}{\bibfnamefont{K.}~\bibnamefont{Hirata}},
  \bibinfo{journal}{Physica C} \textbf{\bibinfo{volume}{445--448}},
  \bibinfo{pages}{260} (\bibinfo{year}{2006}).

\bibitem[{\citenamefont{Ooi et~al.}(2007{\natexlab{a}})\citenamefont{Ooi,
  Mochiku, Gaifullin, and Hirata}}]{physc:ooi07}
\bibinfo{author}{\bibfnamefont{S.}~\bibnamefont{Ooi}},
  \bibinfo{author}{\bibfnamefont{T.}~\bibnamefont{Mochiku}},
  \bibinfo{author}{\bibfnamefont{M.}~\bibnamefont{Gaifullin}},
  \bibnamefont{and} \bibinfo{author}{\bibfnamefont{K.}~\bibnamefont{Hirata}},
  \bibinfo{journal}{Physica C} \textbf{\bibinfo{volume}{460--462}},
  \bibinfo{pages}{1220} (\bibinfo{year}{2007}{\natexlab{a}}).

\bibitem[{\citenamefont{Ooi et~al.}(2007{\natexlab{b}})\citenamefont{Ooi,
  Mochiku, and Hirata}}]{physc:ooi07b}
\bibinfo{author}{\bibfnamefont{S.}~\bibnamefont{Ooi}},
  \bibinfo{author}{\bibfnamefont{T.}~\bibnamefont{Mochiku}}, \bibnamefont{and}
  \bibinfo{author}{\bibfnamefont{K.}~\bibnamefont{Hirata}},
  \bibinfo{journal}{Physica C} \textbf{\bibinfo{volume}{463--465}},
  \bibinfo{pages}{271} (\bibinfo{year}{2007}{\natexlab{b}}).

\bibitem[{\citenamefont{Ooi et~al.}(2005)\citenamefont{Ooi, Mochiku, Yu, Sadki,
  and Hirata}}]{physc:ooi05}
\bibinfo{author}{\bibfnamefont{S.}~\bibnamefont{Ooi}},
  \bibinfo{author}{\bibfnamefont{T.}~\bibnamefont{Mochiku}},
  \bibinfo{author}{\bibfnamefont{S.}~\bibnamefont{Yu}},
  \bibinfo{author}{\bibfnamefont{E.}~\bibnamefont{Sadki}}, \bibnamefont{and}
  \bibinfo{author}{\bibfnamefont{K.}~\bibnamefont{Hirata}},
  \bibinfo{journal}{Physica C} \textbf{\bibinfo{volume}{426--431}},
  \bibinfo{pages}{113} (\bibinfo{year}{2005}).

\bibitem[{\citenamefont{Zeldov et~al.}(1995)\citenamefont{Zeldov, Majer,
  Konczykowski, Geshkenbein, Vinokur, and Shtrikman}}]{nat:zel95}
\bibinfo{author}{\bibfnamefont{E.}~\bibnamefont{Zeldov}},
  \bibinfo{author}{\bibfnamefont{D.}~\bibnamefont{Majer}},
  \bibinfo{author}{\bibfnamefont{M.}~\bibnamefont{Konczykowski}},
  \bibinfo{author}{\bibfnamefont{V.~B.} \bibnamefont{Geshkenbein}},
  \bibinfo{author}{\bibfnamefont{V.~M.} \bibnamefont{Vinokur}},
  \bibnamefont{and}
  \bibinfo{author}{\bibfnamefont{H.}~\bibnamefont{Shtrikman}},
  \bibinfo{journal}{Nature} \textbf{\bibinfo{volume}{375}},
  \bibinfo{pages}{373} (\bibinfo{year}{1995}).

\bibitem[{\citenamefont{Soibel et~al.}(2000)\citenamefont{Soibel, Zeldov,
  Rappaport, Myasoedov, Tamegai, Ooi, Konczykowski, and
  Geshkenbein}}]{nat:soi00}
\bibinfo{author}{\bibfnamefont{A.}~\bibnamefont{Soibel}},
  \bibinfo{author}{\bibfnamefont{E.}~\bibnamefont{Zeldov}},
  \bibinfo{author}{\bibfnamefont{M.}~\bibnamefont{Rappaport}},
  \bibinfo{author}{\bibfnamefont{Y.}~\bibnamefont{Myasoedov}},
  \bibinfo{author}{\bibfnamefont{T.}~\bibnamefont{Tamegai}},
  \bibinfo{author}{\bibfnamefont{S.}~\bibnamefont{Ooi}},
  \bibinfo{author}{\bibfnamefont{M.}~\bibnamefont{Konczykowski}},
  \bibnamefont{and} \bibinfo{author}{\bibfnamefont{V.~B.}
  \bibnamefont{Geshkenbein}}, \bibinfo{journal}{Nature}
  \textbf{\bibinfo{volume}{406}}, \bibinfo{pages}{282} (\bibinfo{year}{2000}).

\bibitem[{\citenamefont{Tokunaga et~al.}(2002)\citenamefont{Tokunaga,
  Kobayashi, Tokunaga, and Tamegai}}]{prb:tok02}
\bibinfo{author}{\bibfnamefont{M.}~\bibnamefont{Tokunaga}},
  \bibinfo{author}{\bibfnamefont{M.}~\bibnamefont{Kobayashi}},
  \bibinfo{author}{\bibfnamefont{Y.}~\bibnamefont{Tokunaga}}, \bibnamefont{and}
  \bibinfo{author}{\bibfnamefont{T.}~\bibnamefont{Tamegai}},
  \bibinfo{journal}{Phys. Rev. B} \textbf{\bibinfo{volume}{66}},
  \bibinfo{pages}{060507(R)} (\bibinfo{year}{2002}).

\bibitem[{\citenamefont{Avraham et~al.}(2008)\citenamefont{Avraham, Brandt,
  Mikitik, Myasoedov, Rappaport, Zeldov, van~der Beek, Konczykowski, and
  Tamegai}}]{prb:avr08}
\bibinfo{author}{\bibfnamefont{N.}~\bibnamefont{Avraham}},
  \bibinfo{author}{\bibfnamefont{E.~H.} \bibnamefont{Brandt}},
  \bibinfo{author}{\bibfnamefont{G.~P.} \bibnamefont{Mikitik}},
  \bibinfo{author}{\bibfnamefont{Y.}~\bibnamefont{Myasoedov}},
  \bibinfo{author}{\bibfnamefont{M.}~\bibnamefont{Rappaport}},
  \bibinfo{author}{\bibfnamefont{E.}~\bibnamefont{Zeldov}},
  \bibinfo{author}{\bibfnamefont{C.~J.} \bibnamefont{van~der Beek}},
  \bibinfo{author}{\bibfnamefont{M.}~\bibnamefont{Konczykowski}},
  \bibnamefont{and} \bibinfo{author}{\bibfnamefont{T.}~\bibnamefont{Tamegai}},
  \bibinfo{journal}{Phys. Rev. B} \textbf{\bibinfo{volume}{77}},
  \bibinfo{eid}{214525} (\bibinfo{year}{2008}).

\bibitem[{\citenamefont{Willemin et~al.}(1998)\citenamefont{Willemin,
  Schilling, Keller, Rossel, Hofer, Welp, Kwok, Olsson, and
  Crabtree}}]{prl:wil98}
\bibinfo{author}{\bibfnamefont{M.}~\bibnamefont{Willemin}},
  \bibinfo{author}{\bibfnamefont{A.}~\bibnamefont{Schilling}},
  \bibinfo{author}{\bibfnamefont{H.}~\bibnamefont{Keller}},
  \bibinfo{author}{\bibfnamefont{C.}~\bibnamefont{Rossel}},
  \bibinfo{author}{\bibfnamefont{J.}~\bibnamefont{Hofer}},
  \bibinfo{author}{\bibfnamefont{U.}~\bibnamefont{Welp}},
  \bibinfo{author}{\bibfnamefont{W.~K.} \bibnamefont{Kwok}},
  \bibinfo{author}{\bibfnamefont{R.~J.} \bibnamefont{Olsson}},
  \bibnamefont{and} \bibinfo{author}{\bibfnamefont{G.~W.}
  \bibnamefont{Crabtree}}, \bibinfo{journal}{Phys. Rev. Lett.}
  \textbf{\bibinfo{volume}{81}}, \bibinfo{pages}{4236} (\bibinfo{year}{1998}).

\bibitem[{\citenamefont{Schmidt et~al.}(1997)\citenamefont{Schmidt,
  Konczykowski, Morozov, and Zeldov}}]{prb:sch97}
\bibinfo{author}{\bibfnamefont{B.}~\bibnamefont{Schmidt}},
  \bibinfo{author}{\bibfnamefont{M.}~\bibnamefont{Konczykowski}},
  \bibinfo{author}{\bibfnamefont{N.}~\bibnamefont{Morozov}}, \bibnamefont{and}
  \bibinfo{author}{\bibfnamefont{E.}~\bibnamefont{Zeldov}},
  \bibinfo{journal}{Phys. Rev. B} \textbf{\bibinfo{volume}{55}},
  \bibinfo{pages}{R8705} (\bibinfo{year}{1997}).

\bibitem[{\citenamefont{Morozov
  et~al.}(1996{\natexlab{a}})\citenamefont{Morozov, Zeldov, Majer, and
  Konczykowski}}]{prb:mor96}
\bibinfo{author}{\bibfnamefont{N.}~\bibnamefont{Morozov}},
  \bibinfo{author}{\bibfnamefont{E.}~\bibnamefont{Zeldov}},
  \bibinfo{author}{\bibfnamefont{D.}~\bibnamefont{Majer}}, \bibnamefont{and}
  \bibinfo{author}{\bibfnamefont{M.}~\bibnamefont{Konczykowski}},
  \bibinfo{journal}{Phys. Rev. B} \textbf{\bibinfo{volume}{54}},
  \bibinfo{pages}{R3784} (\bibinfo{year}{1996}{\natexlab{a}}).

\bibitem[{\citenamefont{Wengel and T\"auber}(1998)}]{prb:wen98}
\bibinfo{author}{\bibfnamefont{C.}~\bibnamefont{Wengel}} \bibnamefont{and}
  \bibinfo{author}{\bibfnamefont{U.~C.} \bibnamefont{T\"auber}},
  \bibinfo{journal}{Phys. Rev. B} \textbf{\bibinfo{volume}{58}},
  \bibinfo{pages}{6565} (\bibinfo{year}{1998}).

\bibitem[{\citenamefont{Majer et~al.}(1995)\citenamefont{Majer, Zeldov, and
  Konczykowski}}]{prl:maj95}
\bibinfo{author}{\bibfnamefont{D.}~\bibnamefont{Majer}},
  \bibinfo{author}{\bibfnamefont{E.}~\bibnamefont{Zeldov}}, \bibnamefont{and}
  \bibinfo{author}{\bibfnamefont{M.}~\bibnamefont{Konczykowski}},
  \bibinfo{journal}{Phys. Rev. Lett.} \textbf{\bibinfo{volume}{75}},
  \bibinfo{pages}{1166} (\bibinfo{year}{1995}).

\bibitem[{\citenamefont{Avraham et~al.}(2001)\citenamefont{Avraham, Khaykovich,
  Myasoedov, Rappaport, Shtrikman, Feldman, Tamegai, Kes, Li, Konczykowski
  et~al.}}]{nat:avr01}
\bibinfo{author}{\bibfnamefont{N.}~\bibnamefont{Avraham}},
  \bibinfo{author}{\bibfnamefont{B.}~\bibnamefont{Khaykovich}},
  \bibinfo{author}{\bibfnamefont{Y.}~\bibnamefont{Myasoedov}},
  \bibinfo{author}{\bibfnamefont{M.}~\bibnamefont{Rappaport}},
  \bibinfo{author}{\bibfnamefont{H.}~\bibnamefont{Shtrikman}},
  \bibinfo{author}{\bibfnamefont{D.~E.} \bibnamefont{Feldman}},
  \bibinfo{author}{\bibfnamefont{T.}~\bibnamefont{Tamegai}},
  \bibinfo{author}{\bibfnamefont{P.~H.} \bibnamefont{Kes}},
  \bibinfo{author}{\bibfnamefont{M.}~\bibnamefont{Li}},
  \bibinfo{author}{\bibfnamefont{M.}~\bibnamefont{Konczykowski}},
  \bibnamefont{et~al.}, \bibinfo{journal}{Nature}
  \textbf{\bibinfo{volume}{411}}, \bibinfo{pages}{451} (\bibinfo{year}{2001}).

\bibitem[{\citenamefont{Beidenkopf et~al.}(2005)\citenamefont{Beidenkopf,
  Avraham, Myasoedov, Shtrikman, Zeldov, Rosenstein, Brandt, and
  Tamegai}}]{prl:bei05}
\bibinfo{author}{\bibfnamefont{H.}~\bibnamefont{Beidenkopf}},
  \bibinfo{author}{\bibfnamefont{N.}~\bibnamefont{Avraham}},
  \bibinfo{author}{\bibfnamefont{Y.}~\bibnamefont{Myasoedov}},
  \bibinfo{author}{\bibfnamefont{H.}~\bibnamefont{Shtrikman}},
  \bibinfo{author}{\bibfnamefont{E.}~\bibnamefont{Zeldov}},
  \bibinfo{author}{\bibfnamefont{B.}~\bibnamefont{Rosenstein}},
  \bibinfo{author}{\bibfnamefont{E.~H.} \bibnamefont{Brandt}},
  \bibnamefont{and} \bibinfo{author}{\bibfnamefont{T.}~\bibnamefont{Tamegai}},
  \bibinfo{journal}{Phys. Rev. Lett.} \textbf{\bibinfo{volume}{95}},
  \bibinfo{eid}{257004} (\bibinfo{year}{2005}).

\bibitem[{\citenamefont{Beidenkopf et~al.}(2007)\citenamefont{Beidenkopf,
  Verdene, Myasoedov, Shtrikman, Zeldov, Rosenstein, Li, and
  Tamegai}}]{prl:bei07}
\bibinfo{author}{\bibfnamefont{H.}~\bibnamefont{Beidenkopf}},
  \bibinfo{author}{\bibfnamefont{T.}~\bibnamefont{Verdene}},
  \bibinfo{author}{\bibfnamefont{Y.}~\bibnamefont{Myasoedov}},
  \bibinfo{author}{\bibfnamefont{H.}~\bibnamefont{Shtrikman}},
  \bibinfo{author}{\bibfnamefont{E.}~\bibnamefont{Zeldov}},
  \bibinfo{author}{\bibfnamefont{B.}~\bibnamefont{Rosenstein}},
  \bibinfo{author}{\bibfnamefont{D.}~\bibnamefont{Li}}, \bibnamefont{and}
  \bibinfo{author}{\bibfnamefont{T.}~\bibnamefont{Tamegai}},
  \bibinfo{journal}{Phys. Rev. Lett.} \textbf{\bibinfo{volume}{98}},
  \bibinfo{eid}{167004} (\bibinfo{year}{2007}).

\bibitem[{\citenamefont{Morozov
  et~al.}(1996{\natexlab{b}})\citenamefont{Morozov, Zeldov, Majer, and
  Khaykovich}}]{prl:mor96}
\bibinfo{author}{\bibfnamefont{N.}~\bibnamefont{Morozov}},
  \bibinfo{author}{\bibfnamefont{E.}~\bibnamefont{Zeldov}},
  \bibinfo{author}{\bibfnamefont{D.}~\bibnamefont{Majer}}, \bibnamefont{and}
  \bibinfo{author}{\bibfnamefont{B.}~\bibnamefont{Khaykovich}},
  \bibinfo{journal}{Phys. Rev. Lett.} \textbf{\bibinfo{volume}{76}},
  \bibinfo{pages}{138} (\bibinfo{year}{1996}{\natexlab{b}}).

\bibitem[{\citenamefont{Schuster et~al.}(1994)\citenamefont{Schuster, Indenbom,
  Kuhn, Brandt, and Konczykowski}}]{prl:sch94}
\bibinfo{author}{\bibfnamefont{T.}~\bibnamefont{Schuster}},
  \bibinfo{author}{\bibfnamefont{M.~V.} \bibnamefont{Indenbom}},
  \bibinfo{author}{\bibfnamefont{H.}~\bibnamefont{Kuhn}},
  \bibinfo{author}{\bibfnamefont{E.~H.} \bibnamefont{Brandt}},
  \bibnamefont{and}
  \bibinfo{author}{\bibfnamefont{M.}~\bibnamefont{Konczykowski}},
  \bibinfo{journal}{Phys. Rev. Lett.} \textbf{\bibinfo{volume}{73}},
  \bibinfo{pages}{1424} (\bibinfo{year}{1994}).

\bibitem[{\citenamefont{Zeldov et~al.}(1994)\citenamefont{Zeldov, Larkin,
  Geshkenbein, Konczykowski, Majer, Khaykovich, Vinokur, and
  Shtrikman}}]{prl:zel94}
\bibinfo{author}{\bibfnamefont{E.}~\bibnamefont{Zeldov}},
  \bibinfo{author}{\bibfnamefont{A.~I.} \bibnamefont{Larkin}},
  \bibinfo{author}{\bibfnamefont{V.~B.} \bibnamefont{Geshkenbein}},
  \bibinfo{author}{\bibfnamefont{M.}~\bibnamefont{Konczykowski}},
  \bibinfo{author}{\bibfnamefont{D.}~\bibnamefont{Majer}},
  \bibinfo{author}{\bibfnamefont{B.}~\bibnamefont{Khaykovich}},
  \bibinfo{author}{\bibfnamefont{V.~M.} \bibnamefont{Vinokur}},
  \bibnamefont{and}
  \bibinfo{author}{\bibfnamefont{H.}~\bibnamefont{Shtrikman}},
  \bibinfo{journal}{Phys. Rev. Lett.} \textbf{\bibinfo{volume}{73}},
  \bibinfo{pages}{1428} (\bibinfo{year}{1994}).

\bibitem[{\citenamefont{Morozov et~al.}(1997)\citenamefont{Morozov, Zeldov,
  Konczykowski, and Doyle}}]{physc:mor97}
\bibinfo{author}{\bibfnamefont{N.}~\bibnamefont{Morozov}},
  \bibinfo{author}{\bibfnamefont{E.}~\bibnamefont{Zeldov}},
  \bibinfo{author}{\bibfnamefont{M.}~\bibnamefont{Konczykowski}},
  \bibnamefont{and} \bibinfo{author}{\bibfnamefont{R.~A.} \bibnamefont{Doyle}},
  \bibinfo{journal}{Physica C} \textbf{\bibinfo{volume}{291}},
  \bibinfo{pages}{113} (\bibinfo{year}{1997}).

\bibitem[{\citenamefont{Stoll et~al.}(2002)\citenamefont{Stoll, Montero,
  Guimpel, \AA{}kerman, and Schuller}}]{prb:sto02}
\bibinfo{author}{\bibfnamefont{O.~M.} \bibnamefont{Stoll}},
  \bibinfo{author}{\bibfnamefont{M.~I.} \bibnamefont{Montero}},
  \bibinfo{author}{\bibfnamefont{J.}~\bibnamefont{Guimpel}},
  \bibinfo{author}{\bibfnamefont{J.~J.} \bibnamefont{\AA{}kerman}},
  \bibnamefont{and} \bibinfo{author}{\bibfnamefont{I.~K.}
  \bibnamefont{Schuller}}, \bibinfo{journal}{Phys. Rev. B}
  \textbf{\bibinfo{volume}{65}}, \bibinfo{pages}{104518}
  (\bibinfo{year}{2002}).

\bibitem[{\citenamefont{Mkrtchyan and Schmidt}(1972)}]{jetp:mkr72}
\bibinfo{author}{\bibfnamefont{G.~S.} \bibnamefont{Mkrtchyan}}
  \bibnamefont{and} \bibinfo{author}{\bibfnamefont{V.~V.}
  \bibnamefont{Schmidt}}, \bibinfo{journal}{Sov. Phys. JETP}
  \textbf{\bibinfo{volume}{34}}, \bibinfo{pages}{195} (\bibinfo{year}{1972}).

\bibitem[{\citenamefont{Wengel and T\"auber}(1997)}]{prl:wen97}
\bibinfo{author}{\bibfnamefont{C.}~\bibnamefont{Wengel}} \bibnamefont{and}
  \bibinfo{author}{\bibfnamefont{U.~C.} \bibnamefont{T\"auber}},
  \bibinfo{journal}{Phys. Rev. Lett.} \textbf{\bibinfo{volume}{78}},
  \bibinfo{pages}{4845} (\bibinfo{year}{1997}).

\bibitem[{\citenamefont{Buzdin}(1993)}]{prb:buz93}
\bibinfo{author}{\bibfnamefont{A.~I.} \bibnamefont{Buzdin}},
  \bibinfo{journal}{Phys. Rev. B} \textbf{\bibinfo{volume}{47}},
  \bibinfo{pages}{11416} (\bibinfo{year}{1993}).

\bibitem[{\citenamefont{Jacobs et~al.}(1995)\citenamefont{Jacobs, Sridhar, Li,
  Gu, and Koshizuka}}]{prl:jac95}
\bibinfo{author}{\bibfnamefont{T.}~\bibnamefont{Jacobs}},
  \bibinfo{author}{\bibfnamefont{S.}~\bibnamefont{Sridhar}},
  \bibinfo{author}{\bibfnamefont{Q.}~\bibnamefont{Li}},
  \bibinfo{author}{\bibfnamefont{G.~D.} \bibnamefont{Gu}}, \bibnamefont{and}
  \bibinfo{author}{\bibfnamefont{N.}~\bibnamefont{Koshizuka}},
  \bibinfo{journal}{Phys. Rev. Lett.} \textbf{\bibinfo{volume}{75}},
  \bibinfo{pages}{4516} (\bibinfo{year}{1995}).

\bibitem[{\citenamefont{Bezryadin et~al.}(1996)\citenamefont{Bezryadin,
  Ovchinnikov, and Pannetier}}]{prb:bez96}
\bibinfo{author}{\bibfnamefont{A.}~\bibnamefont{Bezryadin}},
  \bibinfo{author}{\bibfnamefont{Y.~N.} \bibnamefont{Ovchinnikov}},
  \bibnamefont{and}
  \bibinfo{author}{\bibfnamefont{B.}~\bibnamefont{Pannetier}},
  \bibinfo{journal}{Phys. Rev. B} \textbf{\bibinfo{volume}{53}},
  \bibinfo{pages}{8553} (\bibinfo{year}{1996}).

\bibitem[{\citenamefont{Raedts et~al.}(2004)\citenamefont{Raedts, Silhanek,
  Van~Bael, and Moshchalkov}}]{prb:rae04}
\bibinfo{author}{\bibfnamefont{S.}~\bibnamefont{Raedts}},
  \bibinfo{author}{\bibfnamefont{A.~V.} \bibnamefont{Silhanek}},
  \bibinfo{author}{\bibfnamefont{M.~J.} \bibnamefont{Van~Bael}},
  \bibnamefont{and} \bibinfo{author}{\bibfnamefont{V.~V.}
  \bibnamefont{Moshchalkov}}, \bibinfo{journal}{Phys. Rev. B}
  \textbf{\bibinfo{volume}{70}}, \bibinfo{pages}{024509}
  (\bibinfo{year}{2004}).

\bibitem[{\citenamefont{Goldschmidt and Opferman}()}]{mis:goldschmidt}
\bibinfo{author}{\bibfnamefont{Y.~Y.} \bibnamefont{Goldschmidt}}
  \bibnamefont{and} \bibinfo{author}{\bibfnamefont{M.}~\bibnamefont{Opferman}},
  \emph{\bibinfo{title}{unpublished}}.

\bibitem[{\citenamefont{Fasano et~al.}(2004)\citenamefont{Fasano, Menghini, and
  de~la Cruz}}]{physc:fas04}
\bibinfo{author}{\bibfnamefont{Y.}~\bibnamefont{Fasano}},
  \bibinfo{author}{\bibfnamefont{M.}~\bibnamefont{Menghini}}, \bibnamefont{and}
  \bibinfo{author}{\bibfnamefont{F.}~\bibnamefont{de~la Cruz}},
  \bibinfo{journal}{Physica C} \textbf{\bibinfo{volume}{408--410}},
  \bibinfo{pages}{474} (\bibinfo{year}{2004}).

\bibitem[{\citenamefont{Col et~al.}(2006)\citenamefont{Col, Menon, Geshkenbein,
  and Blatter}}]{prl:dec06}
\bibinfo{author}{\bibfnamefont{A.} \bibnamefont{DeCol}},
  \bibinfo{author}{\bibfnamefont{G.~I.} \bibnamefont{Menon}},
  \bibinfo{author}{\bibfnamefont{V.~B.} \bibnamefont{Geshkenbein}},
  \bibnamefont{and} \bibinfo{author}{\bibfnamefont{G.}~\bibnamefont{Blatter}},
  \bibinfo{journal}{Phys. Rev. Lett.} \textbf{\bibinfo{volume}{96}},
  \bibinfo{eid}{177001} (\bibinfo{year}{2006}).

\end{thebibliography}

\end{document}